\documentclass[aps,pra,reprint,superscriptaddress,nofootinbib,onecolumn,keywords]{revtex4-2}

\usepackage{orcidlink}
\usepackage[T1]{fontenc}
\usepackage[utf8]{inputenc}
\usepackage[english]{babel}
\usepackage{braket}
\usepackage{xcolor}
\usepackage{color}
\usepackage{fancyhdr}
\usepackage{hyperref}
\usepackage{amsfonts, amsmath, amssymb}
\usepackage{graphicx}
\usepackage{comment}
\usepackage{hyperref}
\hypersetup{
	colorlinks = true,
	allcolors = blue,
	pdftitle = {},
}

\newcommand{\mi}{\mathrm{i}}

\begin{document}
	
	\title{Mean-field and fluctuation dynamics in off-resonant two-mode atom–field interactions}
	
	\author{Luis Medina-Dozal\,\orcidlink{0000-0002-4695-5190}}
	\email{luis.medina@icf.unam.mx}
	\affiliation{Instituto de Ciencias F\'isicas, Universidad Nacional Aut\'onoma de M\'exico, Av. Universidad s/n, Col. Chamilpa, 62210, Cuernavaca, Morelos, M\'exico}
	
	\author{Alejandro R. Urzúa\,\orcidlink{0000-0002-6255-5453}}
	\affiliation{Instituto de Ciencias F\'isicas, Universidad Nacional Aut\'onoma de M\'exico, Av. Universidad s/n, Col. Chamilpa, 62210, Cuernavaca, Morelos, M\'exico}
	
	\author{Carlos A. González-Gutiérrez\,\orcidlink{0000-0002-1734-1405}}
	\affiliation{Instituto de Ciencias F\'isicas, Universidad Nacional Aut\'onoma de M\'exico, Av. Universidad s/n, Col. Chamilpa, 62210, Cuernavaca, Morelos, M\'exico}
	
	\author{José Récamier\,\orcidlink{0000-0002-5995-0380}}
	\affiliation{Instituto de Ciencias F\'isicas, Universidad Nacional Aut\'onoma de M\'exico, Av. Universidad s/n, Col. Chamilpa, 62210, Cuernavaca, Morelos, M\'exico}
	
	\begin{abstract}
		We study a two-level system coupled to two quantized electromagnetic modes within the
		Jaynes–Cummings framework. While the single-mode model is exactly solvable due to its
		conserved excitation number, yielding finite-dimensional invariant subspaces, the two-mode model extension presents a fundamental challenge: although the total excitation number remains
		conserved, each invariant subspace is infinite-dimensional, preventing a closed-form analytical
		solution. Our scheme separates the dynamics into a dominant, exactly solvable semiclassical
		component—the atom interacting with the mean fields of both modes—and treats the
		remaining quantum fluctuations through a sequence of unitary transformations that
		preserve essential quantum features. We validate our approach through direct comparison with numerical solutions, focusing on the non-resonant regime where multiple detunings give rise to rich interference effects and multi-timescale dynamics inaccessible to standard approximations. The method accurately reproduces atomic inversion, field observables, and fidelity over relevant timescales, while remaining computationally efficient.
	\end{abstract}
	
	
	\maketitle
	
	\section{Introduction}
	The single-mode Jaynes--Cummings model (JCM) occupies a central position in quantum optics as one of the few interacting light--matter systems whose dynamics admit an exact analytical solution. This solvability rests on two key structural properties. First, the total excitation number is conserved, allowing the total Hilbert space to decompose into invariant two-dimensional subspaces. Second, the operators appearing in the Hamiltonian generate a finite-dimensional dynamical Lie algebra. Together, these features imply that the time-evolution operator can be expressed as a finite product of exponentials with time-dependent coefficients, yielding closed-form solutions for the system’s dynamics
	\cite{Kimble1998,HarocheRaimond2006,WeiNorman1964}. When a two-level system interacts simultaneously with two quantized electromagnetic modes, this algebraic structure is fundamentally altered. Although the total excitation number remains conserved, the operator algebra generated by the interaction no longer closes on a finite set. Mixed-mode commutators produce higher-order operator products involving arbitrarily large powers of the field operators coupled to atomic degrees of freedom. Each successive commutation introduces new operators outside the span of the previous ones, leading to an infinite-dimensional dynamical Lie algebra. As a consequence, the evolution operator of the fully quantized two-mode model cannot be parametrized by a finite number of functions, and the algebraic structure that characterizes the single-mode JCM is lost. From a group-theoretic perspective, this obstruction precludes the direct application of algebraic diagonalization or finite Wei--Norman factorizations
	\cite{WeiNorman1964}.
	
	Despite this increased mathematical complexity, two-mode light--matter interactions arise naturally in a wide range of experimental platforms. In cavity quantum electrodynamics, atoms can couple simultaneously to distinct longitudinal or transverse cavity modes
	\cite{Kimble1998,HarocheRaimond2006}. Trapped-ion systems generate equivalent interactions through bichromatic laser driving that couples internal electronic transitions to different motional sidebands
	\cite{Leibfried2003,BlattRoos2012,vogel,xiao}. In circuit QED architectures, superconducting qubits coupled to multiple microwave resonators offer a highly controllable solid-state implementation
	\cite{Blais2021,Reagor2016}. Related multimode interactions also occur in ultracold atoms inside optical cavities
	\cite{Baumann2010,Gopalakrishnan2009}
	and in nanophotonic platforms with localized electromagnetic modes
	\cite{Thompson2013,Chang2014}.
	These systems access regimes where mode--mode interference, nonlinear dressing, and coherent frequency mixing play a central role, motivating the development of theoretical tools capable of capturing such dynamics.\\
	A simplification is possible when the electromagnetic fields are treated semiclassically replacing quantum field operators by prescribed classical amplitudes. In this regime, the dynamical symmetry group becomes finite-dimensional once again, restoring the possibility of constructing the evolution operator explicitly through standard techniques such as the Magnus expansion or the Wei--Norman theorem
	\cite{Magnus1954}. Although field quantum fluctuations are neglected, the resulting dynamics retain the essential nonlinear structure induced by multiple coherent drives.
	This work presents an approximation method for the fully quantized two-mode Jaynes–Cummings model, focusing on the off-resonant regime.
	
	Our strategy separates the dynamics into a dominant, exactly solvable semiclassical component and a residual part containing quantum fluctuations. The semiclassical Hamiltonian—describing a quantum two-level atom driven by the mean fields of both modes—serves as the dominant contribution. We move to the interaction picture defined by this Hamiltonian and treat the remaining part through a sequence of unitary transformations that preserve essential quantum features such as conditional dynamics and entanglement.
	
	To validate the method, we compare dynamical observables—atomic inversion, average photon number, and fidelity—with purely numerical simulations of the two-mode JCM. The results show that our approach captures quantum correlations inaccessible to standard semiclassical treatments while remaining computationally efficient, offering a practical alternative to numerical diagonalization in regimes where the latter becomes expensive.
	
	The article is organized as follows. Section~\ref{sec 2} introduces the two-mode Jaynes–Cummings Hamiltonian and discusses the challenges in obtaining an analytical solution. Section~\ref{sec 3} presents the semiclassical reference Hamiltonian, its validity conditions, and establishes the decomposition strategy. Section~\ref{sec 4} solves the semiclassical off-resonant atomic Hamiltonian using two complementary approaches—a first-order Magnus expansion and an exact algebraic solution via the Wei-Norman theorem—and verifies that it accurately reproduces the energy transfer dynamics. Section~\ref{sec 5} treats the residual part, derives the complete approximate solution, and tests the validity of the approximation against purely numerical simulations of the two-mode JCM. Finally, section \ref{sec 6} contains our conclusions.
	
	\section{Model and fully quantized Hamiltonian}\label{sec 2} 
	Consider an atom described as a two-level system (TLS) with a ground state $\ket{g}$ and an excited state $\ket{e}$, separated by a transition frequency $\omega_0$. The atom interacts with two independent electromagnetic modes of frequencies $\omega_1$ and $\omega_2$, described by the canonical Jaynes–Cummings model (JCM) under the rotating wave approximation (RWA). We denote by $\hat{a}_i$ and $\hat{a}_i^\dagger$ the usual bosonic ladder operators for each mode, satisfying $[\hat{a}_i,\hat{a}_j^\dagger] = \delta_{ij}$ and $[\hat{a}_i,\hat{a}_j] = [\hat{a}_i^{\dagger},\hat{a}_j^{\dagger}] = 0$. The TLS is characterized by the Pauli operators $\hat{\sigma}_{z} = \ket{e} \! \bra{e} - \ket{g} \! \bra{g}$, $\hat{\sigma}_{+} = \ket{e} \! \bra{g}$ and$\hat{\sigma}_{-} = \ket{g} \! \bra{e}$ which satisfy the $\mathfrak{su}(2)$ commutation relations $[\hat{\sigma}_{+}, \hat{\sigma}_{-}] = \hat{\sigma}_{z}$ and $[\hat{\sigma}_{z}, \hat{\sigma}_{\pm}] = \pm 2\hat{\sigma}_{\pm}$.
	
	The free evolution part of the JCM  Hamiltonian is (using $\hbar=1$)
	\begin{equation}\label{eq:H0}
		\hat{H}_{0} = \sum_{i = 1}^{2} \omega_{i} \hat{a}_{i}^{\dagger} \hat{a}_{i} + \frac{\omega_{0}}{2} \hat{\sigma}_{z},
	\end{equation}
	and the minimal-coupling (dipole) atom--field interaction is written as
	\begin{equation}\label{eq:Vint}
		\hat{V} = \sum_{i = 1}^2 g_{i} \left( \hat{a}_{i}^{\dagger} \hat{\sigma}_{-} + \hat{a}_{i} \hat{\sigma}_{+} \right),
	\end{equation}
	where $g_i$ are real and positive coupling constants (any phases can be absorbed into $\hat{a}_{i}$ or $\hat{\sigma}_{\pm}$). The full two-mode JCM Hamiltonian is therefore
	\begin{equation}\label{eq:JCM}
		\begin{aligned}
			\hat{H} &= \hat{H}_{0} + \hat{V}\\ 
			&= \sum_{i = 1}^{2} \omega_{i}\, \hat{a}_{i}^{\dagger} \hat{a}_{i} + \frac{\omega_{0}}{2}\, \hat{\sigma}_{z} + \sum_{i = 1}^{2} g_{i}\left(\hat{a}_{i}^{\dagger}\hat\sigma_{-} + \hat{a}_{i}\hat{\sigma}_{+} \right).
		\end{aligned}
	\end{equation}
	
	The time evolution operator factorizes as  $\hat{U}(t) = \hat{U}_0(t) \hat{U}_I(t)$ with, 
	\begin{equation}\label{eq;U_0}
		\hat{U}_{0}(t) = e^{-\mi\hat{H}_{0}t},
	\end{equation}
	and $\hat{U}_{I}(t)$ satisfying 
	\begin{equation}
		\mi\partial_{t} \hat{U}_{I}(t) = \left(\hat{U}_{0}^{\dagger}(t)\hat{V}\hat{U}_{0}(t)\right) \hat{U}_{I}(t) = \hat{H}_{I}(t) \hat{U}_{I}(t),
	\end{equation}
	$\hat{U}_I(t)$ is subject to the initial condition $\hat{U}_I(t_0)=\cal{I}$. Thus, the interaction picture Hamiltonian is
	\begin{equation}\label{eq:H_I}
		\hat{H}_{I}(t) = \sum_{i = 1}^{2} g_{i}\left(\hat{a}_{i}^{\dagger}\hat{\sigma}_{-} e^{-\mi\Delta_{i} t} + \hat{a}_{i} \hat{\sigma}_{+} e^{\mi\Delta_{i} t} \right),
	\end{equation}
	where $\Delta_i = \omega_0 - \omega_i$ are the detunings between the atomic transition and each cavity mode. From a Lie–algebraic perspective, the difficulty in solving the fully quantized two-mode JCM stems from the fact that the interaction-picture Hamiltonian in~\eqref{eq:H_I} does not generate a finite-dimensional dynamical Lie algebra.
	
	In the one-mode JCM, the conservation of the total excitation number $\hat{N} = \hat{a}^{\dagger}\hat{a} + \tfrac{1}{2}(\hat{\sigma}_{z} + 1)$ induces a decomposition of the Hilbert space into invariant two-dimensional subspaces spanned by $\{\ket{e, n}, \ket{g, n + 1}\}$. Within each sector, the dynamics reduce to that of an effective two-level system, allowing a direct diagonalization in terms of $2\times 2$ blocks~\cite{HarocheRaimond2006}. From a group-theoretical perspective, this reduction is reflected in the existence of a closed dynamical Lie algebra. Indeed, introducing the composite operators $\hat{b} = (1/\sqrt{\hat{N}})\hat{a}\hat{\sigma}_{+}$ and $\hat{b}^{\dagger} = \hat{a}^{\dagger}\hat{\sigma}_{-}(1/\sqrt{\hat{N}})$~\cite{blas-moya,moya, medina2020}, one finds that the set $\{\hat{b}, \hat{b}^{\dagger}, \hat{\sigma}_{z}\}$ closes under commutation and realizes an $\mathfrak{su}(2)$-type algebra, with commutation relations $[\hat{b}, \hat{b}^{\dagger}] = \hat{\sigma}_{z}$, $[\hat{\sigma}_{z}, \hat{b}] = 2\hat{b}$, and $[\hat{\sigma}_{z}, \hat{b}^{\dagger}] = -2\hat{b}^{\dagger}$. Equivalently, starting from the operator set $\mathcal{S}_{1} = \{\hat{\sigma}_{z}, \hat{n}, \hat{a}\hat{\sigma}_{+}, \hat{a}^{\dagger}\hat{\sigma}_{-}\}$, repeated commutation does not generate independent field or spin operators. Still, it remains confined to composite operators such as $\hat{n}\hat{\sigma}_{z}$ within the same algebraic span. This closed structure can be equivalently represented in the operator basis $\{\hat{a}, \hat{a}^{\dagger}, \hat{\sigma}_{\pm},\hat{\sigma}_{z}\}$, which spans the semidirect sum $\mathfrak{h}_{4} \rtimes \mathfrak{su}(2)$ constrained by $\hat{N}$, and ensures that the time-evolution operator can be expressed as a finite product of exponentials via diagonalization or the Wei--Norman construction~\cite{WeiNorman1964,recamier2018}. On the other hand, in the two-mode extension, although the total excitation number $\hat{N} = \hat{n}_{1} + \hat{n}_{2} + \frac{1}{2}(1 + \hat{\sigma}_{z})$ remains conserved, the algebraic structure is qualitatively different. The interaction involves the set $\{\hat{a}_{1}\hat{\sigma}_{+}, \hat{a}_{2}\hat{\sigma}_{+}, \hat{a}_{1}^{\dagger}\hat{\sigma}_{-}, \hat{a}_{2}^{\dagger}\hat{\sigma}_{-}\}$, whose commutators generate mixed-mode operators such as $\hat{a}_{1}^{\dagger}\hat{a}_{1}\hat{\sigma}_{z}$, $\hat{a}_{2}^\dagger\hat{a}_{2}\hat{\sigma}_{z}$, $\hat{a}_{1}\hat{a}_{2}^\dagger\hat{\sigma}_{z}$, and $\hat{a}_{1}^\dagger\hat{a}_{2}\hat{\sigma}_{z}$, together with projectors $\ket{e}\bra{e}$ and $\ket{g}\bra{g}$. As a result, the dynamical Lie algebra does not close on a finite set, but instead proliferates to an infinite-dimensional structure. At the level of invariant subspaces, the fixed $N$ conditions, $n_{1} + n_{2} = N - 1$ (excited state) and $n_{1} + n_{2} = N$ (ground state) admit multiple field configurations. Hence, each sector is no longer two-dimensional but grows with $N$. Consequently, no analog of the $\hat{b}$-operator construction exists, and the evolution operator cannot be factorized in terms of a finite number of algebra generators.
	
	This loss of finite-dimensional dynamical symmetry, not the absence of a conserved quantity, is the essential obstruction to exact solvability in the fully quantized two-mode model. It motivates the semiclassical reduction adopted in the following sections. When the field operators are replaced by their c-number mean amplitudes $\hat{a}_{i} \to \alpha_{i}(t)$, the operator content collapses to the closed algebra $\{\hat{\sigma}_{\pm},\hat{\sigma}_{z}\}\cong\mathfrak{su}(2)$. The resulting semiclassical Hamiltonian is closed under commutation and forms a finite-dimensional Lie algebra, allowing the evolution operator to be constructed explicitly employing the Wei--Norman theorem~\cite{WeiNorman1964}.
	
	\section{Semiclassical Hamiltonian approach}\label{sec 3}
	Given the complexity of the fully quantized problem, which encompasses both the electromagnetic fields and the two-level system, we adopt a semiclassical approach suitable for the parameter regime under consideration. Within this approximation, the field operators in the interaction Hamiltonian are replaced by their expectation values. When the field amplitudes are large enough, specifically when the fields are initially in coherent states with large photon numbers $|\alpha_i|^2 \gg 1$, the substitution $\hat{a}_i\mapsto \langle \hat{a}_i\rangle = \alpha_i$ where $\alpha_i$ is a complex number representing the amplitude of the semiclassical field associated with the coherent state $|\alpha_i\rangle$ is justified.  In this regime, quantum fluctuations are small compared to the mean field amplitude, as they scale as $1/|\alpha_i|$. This replacement suppresses the quantum description of the fields, enabling a tractable treatment of the atomic dynamics.  Within this approximation, the dominant physical process, coherent population transfer between the atom and the fields, can be captured by a much simpler object: a two-level system driven by classical fields. 
	Accordingly, we can express a semiclassical version of the interaction Hamiltonian~\eqref{eq:H_I} as
	\begin{equation}\label{eq:H_sc}
		\hat{H}_{\mathrm{sc,I}}(t) = \sum_{i = 1}^{2} \left[ g_{i}\alpha_{i}e^{\mi\Delta_i t}\, \hat{\sigma}_{+} + g_{i}\alpha_{i}^{*}e^{-\mi\Delta_i t}\, \hat{\sigma}_{-} \right],
	\end{equation}
	where we define the \emph{semiclassical Rabi amplitude}
	\begin{equation}
		\Omega_{\mathrm{sc}}(t) \equiv \sum_{i = 1}^{2} g_{i}\alpha_{i}e^{\mi\Delta_i t},
	\end{equation}
	so that \eqref{eq:H_sc} becomes $\hat H_{\mathrm{sc,I}}(t) =\Omega_{\mathrm{sc}}(t) \hat{\sigma}_{+} + \Omega_{\mathrm{sc}}^{*}(t) \hat{\sigma}_{-}$.
	
	We now perform an exact decomposition of the  interaction Hamiltonian by adding and subtracting the semiclassical Hamiltonian $\hat{H}_{\mathrm{sc,I}}(t)$:
	\begin{equation}\label{eq:decomposition}
		\hat{H}_I(t) = \hat{H}_{\mathrm{sc,I}}(t) + \big(\hat{H}_I(t) - \hat{H}_{\mathrm{sc,I}}(t)\big)\equiv \hat{H}_{sc,I}(t) + \hat{V}^{(1)}(t).
	\end{equation}
	This decomposition splits the problem into a solvable semiclassical part and a residual part. \(\hat{H}_{\mathrm{sc,I}}(t)\) captures the main population dynamics, that is, the coherent energy exchange between the atom and the classical mean fields, while the residual part
	\begin{equation}
		\hat{V}^{(1)}(t)=\sum_i g_ie^{\mi\Delta_it}\hat a_i\hat\sigma_+-\Omega_{sc}(t)\hat\sigma_++\mathrm{h.c.}
	\end{equation}
	contains the quantum fluctuations and correlations beyond this mean-field description. The treatment of $\hat{V}^{(1)}(t)$, which encodes these quantum correlations, will be developed in Sec.~\ref{sec 5}.
	
	\section{Semiclassical atomic dynamics.}\label{sec 4}
	
	The Hamiltonian $\hat{H}_{\mathrm{sc,I}}(t)$ describes a quantum two-level system driven by classical fields. This problem is exactly solvable and has been widely studied in the literature both as a fundamental model of driven quantum systems and for its applications in coherent control and quantum information processing \cite{gerry,castanos,alscher,babelon,vidella}. In this section, we first analyze the resonant case; that is,  where both fields and the atom have the same frequency ($\omega_1 = \omega_2 = \omega_0$). We then turn to the more general and physically interesting non-resonant case, where the frequencies $\omega_1$, $\omega_2$, and $\omega_0$ are all different. For this regime, we obtain the atomic evolution using two complementary methods.\\
	
	\subsection{Resonant case: additive effective drive}
	To obtain the atomic dynamics under $\hat{H}_{\mathrm{sc,I}}(t)$, we derive the equations of motion for the relevant expectation values. Defining $s(t)=\langle\hat{\sigma}_-(t)\rangle$ and $W(t)=\langle\hat{\sigma}_z(t)\rangle$, and taking expectation values of the Heisenberg equations for the atomic operators, we obtain
	\begin{align}
		\dot{s}(t) &= \mi \, \Omega_{\mathrm{sc}}(t)\,W(t),\label{eq:mf_s}\\
		\dot{W}(t) &= -4\,\mathrm{Im}\left[\Omega_{\mathrm{sc}}^{*}(t)\, s(t)\right]. \label{eq:mf_w}
	\end{align}
	
	Consider the resonant scenario where both field modes are tuned exactly to the atomic transition, i.e., $\Delta_{i} =\omega_{0} -\omega_{i} = 0$. In the atomic rotating frame, the semiclassical interaction Hamiltonian \eqref{eq:H_sc} reduces to
	\begin{equation}
		\hat H_{\mathrm{sc,I}} = \Omega_{\mathrm{r,sc}}\hat{\sigma}_{+} + \Omega_{\mathrm{r,sc}}^{*}\, \hat{\sigma}_{-} 
	\end{equation}
	with the \emph{resonant Rabi amplitude} $\Omega_{\mathrm{r,sc}} \equiv g_{1}\alpha_{1} + g_{2}\alpha_{2}.$
	
	For the initial condition where the atom starts in the excited state $\ket{e}$ (implying $W(0) = 1$ and $s(0) = 0$), the exact solutions of the
	Eqs. \eqref{eq:mf_s} and \eqref{eq:mf_w} are
	\begin{equation}\label{eq:res_s_w}
		s(t) = \frac{1}{2}\sin\left(2\,\Omega_{\mathrm{r,sc}}\, t\right),\qquad
		W(t) = \cos\left(2\,\Omega_{\mathrm{r,sc}}\,t\right).
	\end{equation}
	These expressions make the key point immediately evident: \emph{on resonance, the two drives combine additively through $\Omega_{\mathrm{r,sc}} = g_{1}\alpha_{1} + g_{2}\alpha_{2}$ to produce a single effective drive strength}. Physically, two coherent fields resonant with the atom directly add their amplitudes to determine the single-tone Rabi frequency.
	
	\subsection{Non-resonant case: First-order Magnus expansion}\label{semiclassical magnus}
	
	To treat the general case of non-resonant fields, $\Delta_{i} \neq 0$, we first pursue an approximate analytical solution via the first-order Magnus expansion \cite{Magnus1954}. This method yields closed-form expressions for $s(t)$ and $W(t)$ (see Appendix~\ref{Appendix A}), which, although approximate, reveal the dominant frequencies and nonlinear mixing processes inherent in the atom-field interaction.
	
	For the initial condition where the system starts in the excited state $\ket{e}$ and considering the perturbative regime $|g_i\alpha_i/\Delta_i|\ll 1$, a first-order Magnus expansion yields an analytical expression for the atomic inversion as
	\begin{equation}\label{eq: magnus inversion}
		W(t)\approx\cos\omega(t)
	\end{equation}
	It is insightful to analyze  $\omega(t)^{2}$ in detail. Using the definitions of $a(t)$ and $b(t)$ from Appendix~\ref{Appendix A}, their combination returns
	\begin{equation}\label{eq:absum}
		\omega(t)^{2} = \sum\limits_{i = 1}^{2} \frac{\left(g_{i}\alpha_{i}\right)^2}{\Delta_{i}^{2}}\sin^{2}\left(\frac{\Delta_{i}t}{2}\right) + 2\sum\limits_{i<j}^{2}\frac{g_{i}g_{j}\alpha_{i}\alpha_{j}}{\Delta_{i}\Delta_{j}}f_{ij}(t),
	\end{equation}
	where
	\begin{equation}
		f_{ij}(t) = 1 - \cos(\Delta_{i}t) - \cos(\Delta_{j}t) + \cos(\Delta_{ij}t),\quad \Delta_{ij} = \Delta_{i} - \Delta_{j}.
	\end{equation}
	We can see that the first term in Eq. \eqref{eq:absum} contains positive-definite oscillatory envelopes. The cross-term $f_{ij}(t)$ represents the mixing between the two detunings, oscillating at the beat frequency $\vert\Delta_{i} - \Delta_{j}\vert$, and because of the sum range, this term is explicitly $\vert\omega_{1} - \omega_{2}\vert$. 
	To investigate the different contributions of each field to atomic dynamics, we deliberately choose a parameter regime that induces an interaction hierarchy. We place the field labeled by $\omega_{2}$ close to atomic resonance by setting it to  $\Delta_{2} \ll 1$, ensuring strong resonant energy exchange, while placing the field labeled by $\omega_{1}$ far from resonance. This establishes that $\omega_{2}$ field is the primary driver of the population transfer. Moreover, we set the coupling strengths to be comparable, $g_{1} \gtrsim g_{2}$, which ensures that the far-detuned field $\omega_{1}$ still exerts a non-negligible influence on the system. This configuration is designed to probe the interplay between a dominant resonant process and perturbative off-resonant effects, allowing us to isolate the frequency shifts, nonlinear mixing, and dressing phenomena that arise specifically from the simultaneous presence of both fields.
	
\begin{figure}[htbp]
		\centering     
		\includegraphics[width = \linewidth, keepaspectratio]{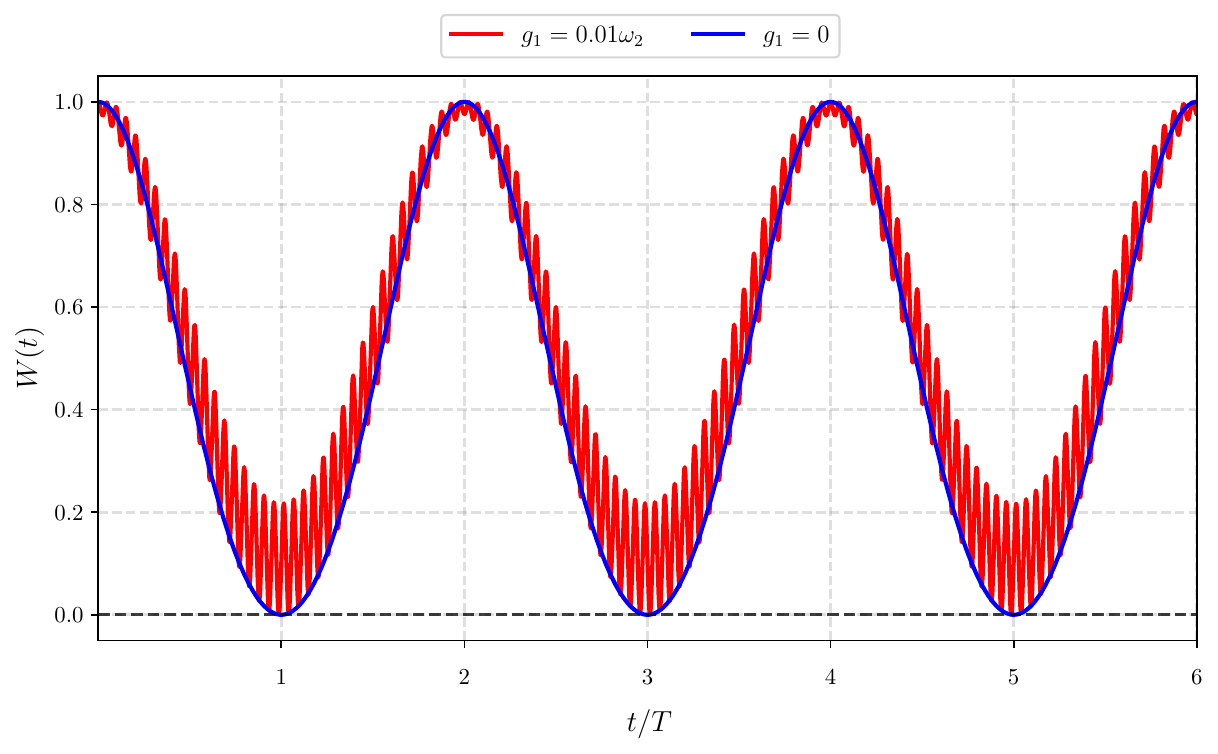}
		\caption{Atomic inversion $W(t)$ using the first order Magnus expansion. Hamiltonian parameters: $\omega_{2} = 1,\omega_{1} = \omega_{2}/2,\omega_{a}= 0.98\omega_{2}, g_{2} = g_{2}^{\texttt{opt}}, \alpha_{1} = \alpha_{2} = 4$. We see the slow frequency envelope when $g_{1} = 0$ (blue) and high frequency wiggles when $g_{1} \neq 0 $ (red). The time scale is given in terms of the period $T = \pi/\Delta_{2}$.}
		\label{fig_01}
	\end{figure}

	Despite the comparable coupling strengths, the significant disparity in detunings leads to a large imbalance in the effective drive amplitudes, $g_{2}\alpha_{2}/\Delta_{2} \gg g_{1}\alpha_{1}/\Delta_{1}$. This condition allows us to focus on the essential physics of population transfer driven by the near-resonant field. In this regime, the dynamics separate into distinct components (see Fig.~\ref{fig_01}), where the near-resonant interaction and fast oscillations from the far-detuned field govern a slow global envelope. The envelope itself follows a generalized Rabi oscillation given by Eq.~\eqref{eq: magnus inversion}, where the effective frequency $\omega(t)$ is now time-dependent due to the integrated action of the fields. 
	
	To isolate and analyze this population transfer, we consider the limiting case where the far-detuned field is decoupled, $g_{1} = 0$. Under this condition, the atomic inversion becomes
	\begin{equation}
		W(t) = \cos\left(\frac{4g_{2}\alpha_{2}}{\Delta_{2}} \sin\left(\frac{\Delta_{2} t}{2}\right)\right),
		\label{eq:W_simplified}
	\end{equation}
	which can be employed to engineer maximum quantum coherence, defined as  $W(T) = 0$ (an equal superposition of atomic states), at a specific time  $T$. Solving $W(T) = 0$ yields
	\begin{equation}
		T = \frac{\pi}{\Delta_{2}},
	\end{equation}
	and the condition for maximum coherence becomes
	\begin{equation}
		\cos\left(\frac{4g_{2}\alpha_{2}}{\Delta_{2}}\right) = 0 \quad \Rightarrow \quad \frac{4g_{2}\alpha_{2}}{\Delta_{2}} = (2n + 1)\frac{\pi}{2},
	\end{equation}
	with $n$ a non-negative integer. This gives a parametric family of coupling strengths
	\begin{equation}
		g_2 = (2n+1)\frac{\pi\Delta_2}{8\alpha_2}.
		\label{eq:g2_design}
	\end{equation}
	
	As we stated above, the validity of the first-order Magnus expansion requires that it operates in a perturbative regime, characterized by $\frac{g_{2}\alpha_{2}}{\Delta_{2}} \ll 1$. To satisfy this constraint, we select the fundamental solution, $n = 0$, which gives the minimal coupling strength
	\begin{equation}
		g_{2}^{\texttt{opt}} = \frac{\pi\Delta_{2}}{8\alpha_2},
		\label{eq:g2_optimal}
	\end{equation}
	
	As this choice ensures the parameter $\frac{g_{2}\alpha_{2}}{\Delta_{2}} = \tfrac{\pi}{8} \approx 0.39$, which remains less than one, thereby preserving the validity of the perturbative approximation. Consequently, by tuning the coupling to $g_{2}^{\mathrm{opt}}$, one can achieve maximum quantum coherence at a predictable time $T = \pi/\Delta_2$ while remaining within the valid operational regime of the approximation. 
	
	In contrast, a similar condition for complete population inversion ($W(T)=-1$) cannot be derived within this framework. Reaching $W(T)=-1$ would require $g_2\alpha_2/\Delta_2$ to exceed the perturbative limit, placing it outside the domain where the first-order Magnus expansion remains accurate.
	
	The atomic inversion in a two-level system driven by two off-resonant fields exhibits coherent oscillations that generalize the standard Rabi dynamics through a time-dependent effective frequency $\omega(t)$. The system’s evolution unfolds on two distinct timescales. On the slow scale, the magnitude of $\omega(t)$ is set by the parameters $g_{i} \alpha_{i} / \Delta_{i}$, which are assumed to satisfy the validity of the approximation. This slow variation governs the global envelope of the population transfer between the two atomic states, effectively determining the overall Rabi cycle. Superimposed on this slow oscillation are fast, small-amplitude oscillations arising from the high-frequency behavior of the variables $a(t)$ and $b(t)$, which oscillate at $\Delta_{1}$ and $\Delta_{2}$. These rapid ``wiggles'' are captured as non-perturbative corrections by the first-order Magnus expansion, revealing nuances of the dynamics that resonant treatments neglect. Altogether, this analytic approach provides a clear physical picture of the multi-timescale behavior, showing how off-resonant driving leads to slow population transfer modulated by subtle, high-frequency features.

	\subsection{Semiclassical Hamiltonian: Exact solution}\label{semiclassical exact}
	The time-evolution operator associated with the semiclassical interaction Hamiltonian Eq.~\eqref{eq:H_sc} can be expressed exactly as a product of exponentials (see Appendix~\ref{Appendix B}),  
	\begin{equation}\label{eq:U_sc}
		\hat U_{\mathrm{sc,I}}(t) = e^{\beta_{+}(t)\hat{\sigma}_{+}}e^{\beta_{-}(t)\hat{\sigma}_{-}} e^{\beta_z (t)\hat{\sigma}_z}.
	\end{equation}
	While the complete time evolution operator associated with the semiclassical Hamiltonian is given by the product
	\begin{equation}
		\hat U_{\mathrm{sc}}(t) =\hat U_{0}(t) \hat U_{\mathrm{sc,I}}(t),
	\end{equation} 
	with $\hat U_{0}(t)$ the time evolution operator corresponding to the free  Hamiltonian~\eqref{eq;U_0}. 
	Once we know the time evolution operator, it is straightforward to compute the state vector at any time
	\begin{eqnarray*}
		\vert\psi(t)\rangle_{\mathrm{sc}} &=&   \hat U_{\mathrm{sc}}(t)\vert\psi(0)\rangle.
	\end{eqnarray*}
	For the initial state $\ket{\psi(0)}=\ket{e,\alpha_1,\alpha_2}$  (atom in the excited state and both fields in coherent states), we obtain
	\begin{equation}\label{eq: atom vector state}
		\vert\psi(t)\rangle_{\mathrm{sc}}=e^{\beta_z(t)}\left(e^{-\mi\frac{\omega_0}{2} t}(1+\beta_+(t)\beta_-(t))\vert e\rangle+e^{\mi\frac{\omega_0}{2} t}\beta_-(t)\vert g\rangle\right)\times\ket{\alpha_1(t),\alpha_2(t)}
	\end{equation}
	where $\alpha_1(t)=\alpha_1e^{-i\omega_1t}$ and  $\alpha_2(t)=\alpha_2e^{-i\omega_2t}$.
	From this vector state, the atomic inversion $W(t)$ and the coherence $s(t)$  are given by
	\begin{gather}\label{eq wei-norman atom}
		W(t) = 1+2\beta_{+}(t)\beta_{-}(t),\\
		s(t) = -\beta_+(t)(1+\beta_+(t)\beta_-(t)).
	\end{gather}
	The time-dependent coefficients $\beta_+(t)$, $\beta_-(t)$, and $\beta_z(t)$ are obtained by numerically integrating the coupled ordinary differential equations given in Appendix~\ref{Appendix B} using a Runge-Kutta method (RK45).
	\begin{figure}[htbp]
		\centering
		\includegraphics[width = \linewidth]{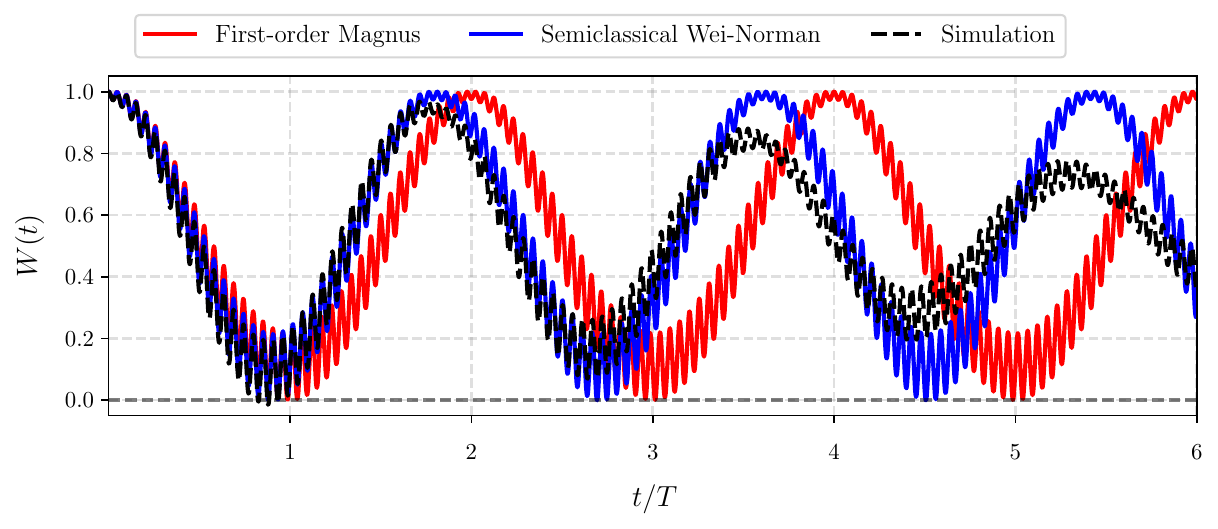}
		\caption{Atomic inversion $W(t)$ using: first order Magnus expansion (red), semiclassical Wei-Norman solution (blue), and numerical simulation (black). Hamiltonian  parameters: $\omega_{2} = 1.0$, $\omega_{0} = 0.98\omega_{2}$, $g_{1} = 0.01\omega_{2}$, $T \approx 157.08$, $\alpha_{1} = \alpha_{2} = 4$, $\omega_{1} = \omega_{2}/4$. In the three cases, $g_{2} = g_{2}^{\texttt{opt}}$ given by \eqref{eq:g2_optimal}.}
		\label{fig_02}
	\end{figure}
	
	In Fig.~\ref{fig_02}, we plot the atomic inversion $W(t)$ obtained from three different methods discussed above. The red line corresponds to the first-order Magnus expansion, the blue line to the Wei-Norman solution of the semiclassical Hamiltonian (Eq.~\eqref{eq wei-norman atom}), and the black line to the numerical solution of the fully quantized two-mode JCM (Eq.~\eqref{eq:JCM}). For all three curves, we use the parameters $\omega_1 = \omega_2/4$, $\omega_2 = 1$, $\omega_0 = 0.98\omega_2$, $g_1 = 0.01\omega_2$, and $g_2 = g_2^{\mathrm{opt}}$ as given by Eq.~\eqref{eq:g2_optimal}.
	The figure shows a slow envelope oscillation, which corresponds to the effective Rabi frequency, with superimposed fast oscillations visible in all three cases. On timescales of the order of the first Rabi cycle, the three solutions agree remarkably well. At later times, however, the Magnus expansion begins to deviate from the other two.
	It is worth noting that both the Magnus expansion and the Wei-Norman solution are derived from the semiclassical Hamiltonian, where the fields are treated as classical drives. Consequently, they predict oscillations of constant amplitude. In contrast, the black curve, obtained from the full simulation, exhibits a gradual decrease in oscillation amplitude at longer times, suggesting convergence toward a steady value. This behavior is a signature of collapses and revivals, a purely quantum feature characteristic of Jaynes–Cummings-type models. 
	\begin{figure}[htbp]
		\centering
		\includegraphics[width = \linewidth]{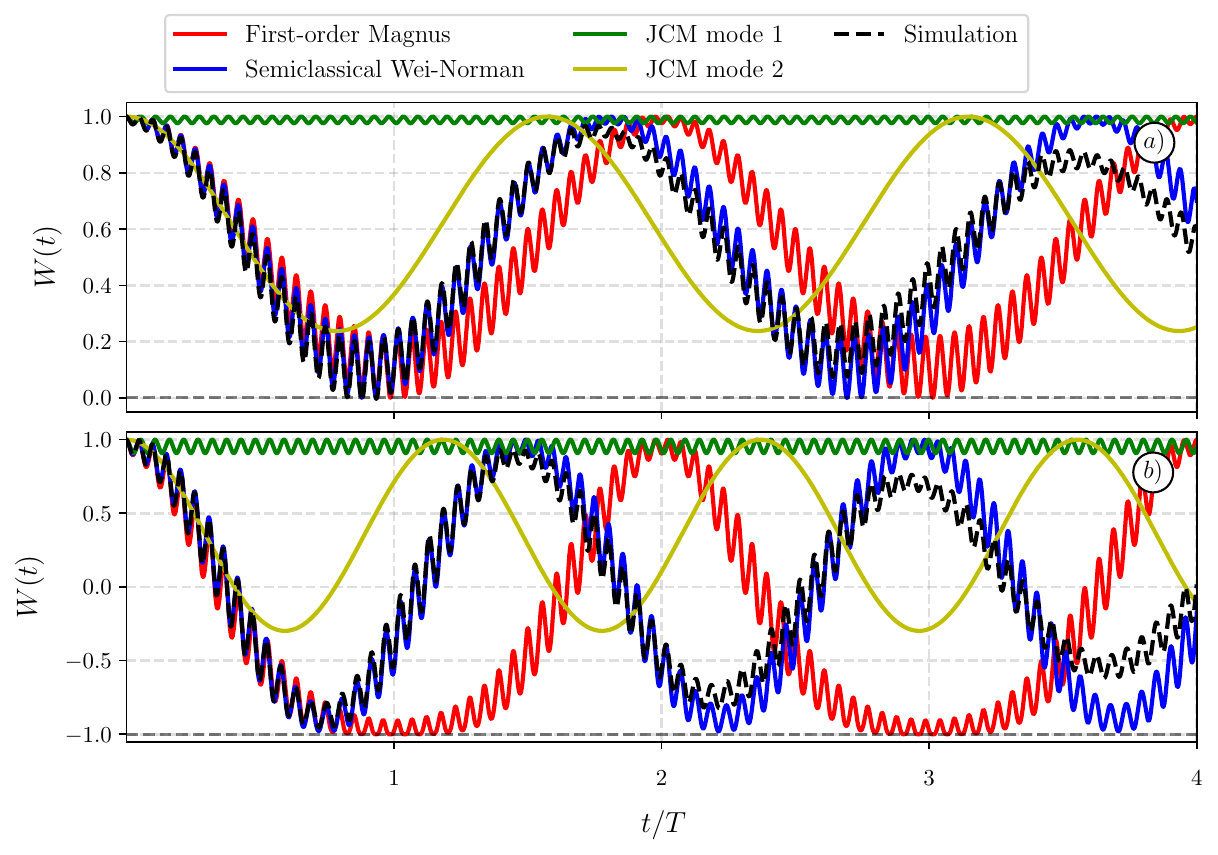}
		\caption{Atomic inversion $W(t)$ using first order Magnus expansion (red), semiclassical Wei-Norman (blue), numerical simulation (black), JCM mode 1 (green), and JCM mode 2 (yellow). Hamiltonian parameters: $\omega_{2} = 1.0$, $\omega_{a} = 0.98\omega_{2}$, $g_{1} = 0.01\omega_{2}$, $T \approx 157.08$, $\omega_{1} = \omega_{2}/4$, $g_{2} = 0.00196\omega_2$ for Magnus, $g_{2} = 0.00213\omega_2$ for Wei-Norman, and $g_2=0.00195\omega_2$ for the simulation. Panel a) $\alpha_{1} = \alpha_{2} = 4$, Panel b) $\alpha_{1} = \alpha_{2} = 8$. }
		\label{fig_03}
	\end{figure}
	Figure~\ref{fig_03} shows the temporal evolution of the atomic inversion. In the upper frame (a), the dark green curve corresponds to the solution obtained by considering only mode 1 of the JCM, i.e., the interaction between the TLS and the electromagnetic field of frequency $\omega_1$ alone. As indicated by the parameters in the caption, $\Delta_1$ is far off resonance. Consequently, the atomic population inversion shows little response to the field mode $\omega_1$. In contrast, the light green curve shows the dynamics of the mode-2 JCM. The amplitude of the atomic inversion oscillations is significantly larger than in the mode-1 case, since mode 2 is much closer to resonance with the TLS.
	At this point, one might be tempted to think that mode 1 acts merely as a passive observer, or at most as a small perturbation on the full two-mode JCM. However, because the coupling strengths are comparable, both modes play a relevant and non-trivial role. This is evident from the red (Magnus), blue (Wei-Norman), and black (full quantum) curves, which show the full two-mode dynamics.
	Panel (b) presents a similar comparison, but in a parameter regime where neither single-mode JCM alone can achieve full atomic inversion. For a single-mode JCM with detuning $\Delta$, the atomic inversion is given by
	\begin{equation*}
		W(t) = \frac{\Delta^2}{\Delta^2 + \Omega_n^2} + \frac{\Omega_n^2}{\Delta^2 + \Omega_n^2} \cos\left( \sqrt{\Delta^2 + \Omega_n^2}\, t \right),
	\end{equation*}
	where $\Omega_n$ is the dressed resonant Rabi frequency. Complete atomic inversion, i.e., $W(t) = -1$, requires $\Delta = 0$.
	In the two-mode case, however, the situation is qualitatively different. Even when both fields are individually off-resonance, their combined effect can drive the atom to full inversion. As revealed by the Magnus expansion in Eq.~\eqref{eq: magnus inversion}, the effective Rabi frequency is no longer a simple function of individual detunings, acquiring a non-trivial structure arising from the interference term $f_{ij}(t)$ in Eq.~\eqref{eq:absum}. This interference allows the two fields to cooperatively overcome the detuning limitations of the single-mode case. The atom thus experiences an effective drive whose strength and frequency depend on both fields simultaneously, enabling the full inversion observed in the red, blue, and black curves of panel (b).
	Taken together, both panels illustrate a central result: while single-mode off-resonant driving is fundamentally limited in its ability to produce large atomic excursions, the cooperative action of two modes, mediated by the interference term $f_{ij}(t)$, can overcome these limitations, leading to maximal coherence and even complete inversion under appropriate conditions. This cooperative enhancement is a genuine two-mode effect that cannot be captured by any single-mode treatment.
	
	It is worth noting a fundamental distinction in obtaining the optimal near-resonant coupling strength, $g_{2}$, across the different methodologies. While the first-order Magnus expansion yields an analytical, closed-form expression to calculate the precise coupling $g_{2}^{(\texttt{opt})}$ required to achieve maximum quantum coherence, this is not the case for the semiclassical Wei-Norman or the exact numerical simulation approaches. To determine the target coupling for these latter methods, we rely on a numerical parameter optimization routine. By setting the objective to reach $\braket{\hat{\sigma}_{z}(T)} = 0$ at the specific target time $T$, we execute a concurrent grid search over a bounded interval $g^{(\texttt{min})}_{2} < g_{2} < g^{(\texttt{max})}_{2}$. Controlling for granular precision, the routine evaluates the resulting population inversions and halts once the target zero-crossing is achieved. The analytical value $g_{2}^{(\texttt{opt})}$ derived from the Magnus method serves as a highly effective initial seed to establish the boundaries of this numerical search.
	
	Given the agreement observed in the atomic inversion dynamics, we now turn to a more stringent test: the ability of the semiclassical approximation to reproduce the time-evolved wavefunction obtained from the full quantum simulation. To quantify this, we compute the quantum fidelity between the reduced atomic states obtained from the semiclassical approximation and from a numerical simulation of the Hamiltonian (Eq.~\eqref{eq:JCM}).
	The fidelity $\mathcal{F}(t)$  between two quantum states 
	$\rho_1(t)$ and $\rho_2(t)$, expressed as density matrices, is commonly defined as \cite{jozsa1994fidelity}
	\begin{equation}
		\mathcal{F}(t) = \left[ \text{Tr} \left( \sqrt{ \sqrt{\hat{\rho}_{1}(t)} \hat{\rho}_{2}(t) \sqrt{\hat{\rho}_{1}(t)}} \right) \right]^2.
		\label{eq:atomic_fidelity}
	\end{equation}
	We define $\hat{\rho}_{\texttt{num}}^{\texttt{(atom)}}(t)$ as the reduced density matrix of the atom obtained by tracing out the field degrees of freedom from the numerical simulation, and $\hat{\rho}_{\texttt{sc}}^{\texttt{(atom)}}(t)$ as the reduced atomic density matrix obtained from the semiclassical product state $\ket{\psi(t)}_{\mathrm{sc}}$ defined in Eq.~\eqref{eq: atom vector state}. The fidelity can be computed by substituting  $\hat{\rho}_{\texttt{num}}^{\texttt{(atom)}}(t)$ and $\hat{\rho}_{\texttt{sc}}^{\texttt{(atom)}}(t)$ in equation~\eqref{eq:atomic_fidelity}. 
	
\begin{figure}[hbtp]
		\centering
		\includegraphics[width = \linewidth]{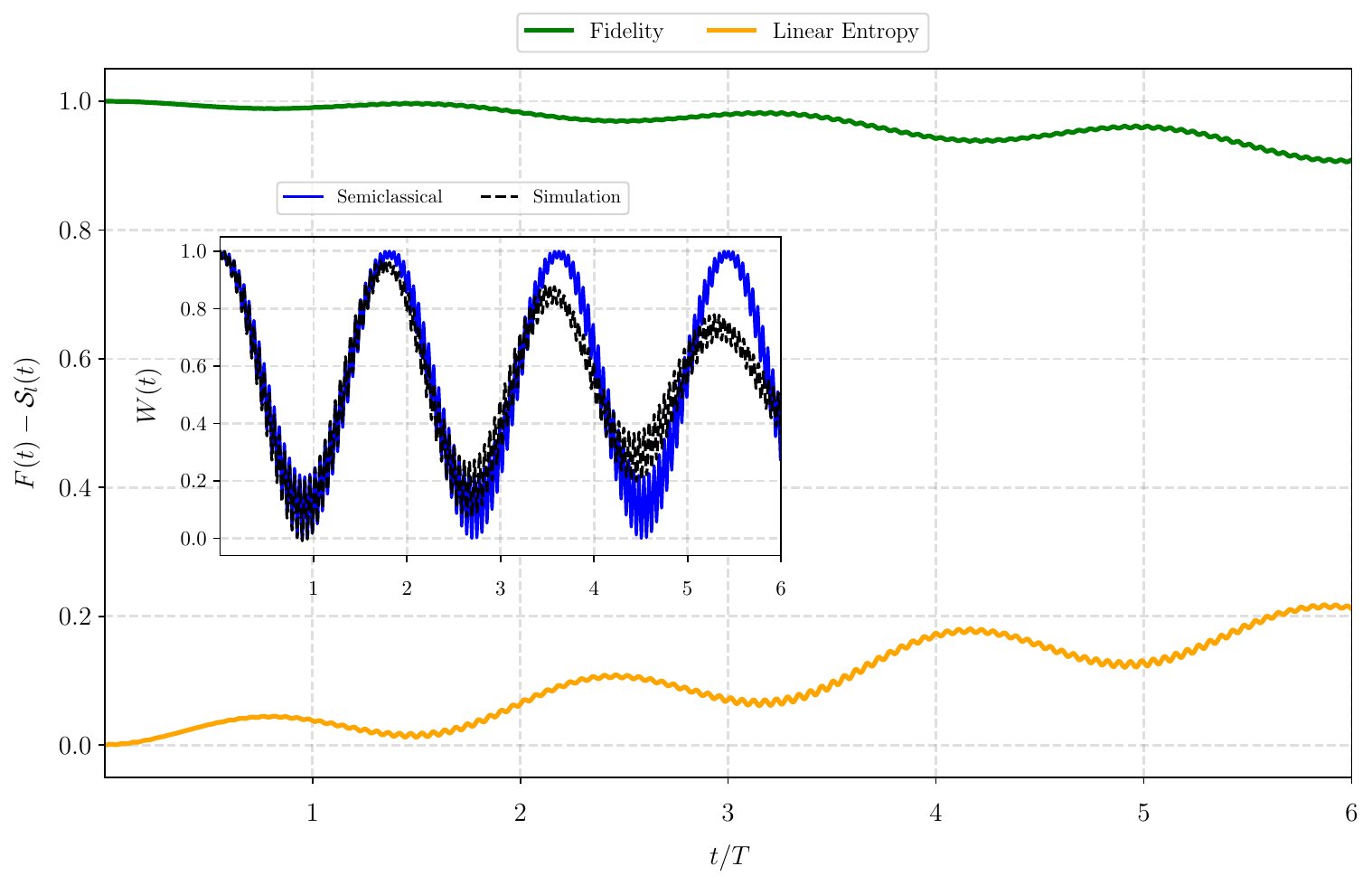}
		\caption{Fidelity  between Wei-Norman and quantum simulation, and linear entropy  of the full quantum simulation. Parameters: $\omega_{2} = 1.0$, $\omega_{0} = 0.98\omega_{2}$, $g_{1} = 0.01\omega_{2}$, $T \approx 157.08$, $\omega_{1} = \omega_{2}/4$, $\alpha_{1} = \alpha_{2} = 4$, $g_{2}$ is the optimal $g_{2}^{(\texttt{opt})}=0.00196\omega_2$. The inset shows the evolution of the population inversions, whereas the main plot displays the fidelity $\mathcal{F}(t)$  in green and the linear entropy $S_{l}(t)$ in orange.}
		\label{fig_04_I}
	\end{figure}

	A fidelity of unity indicates perfect agreement, whereas deviations quantify the error introduced by the semiclassical approximation. Numerical results indicate that the atomic fidelity exhibits a pronounced oscillatory behavior accompanied by a slow decay (see Fig.~\ref{fig_04_I}). To elucidate the physical mechanism driving these deviations, we analyze the entanglement accumulation in the full quantum system. Since the total system evolves unitarily from a pure state, the entropy of the reduced atomic subsystem serves as a direct measure of atom-field entanglement. We utilize the linear entropy $S_{l}(t)$, defined as
	\begin{equation}
		S_{l}(t) = 1 - \mathrm{Tr}\left[ (\hat{\rho}_{\texttt{num}}^{\texttt{(atom)}}(t))^2 \right]
		\label{eq:linear_entropy}
	\end{equation}
	which ranges from $0$ (for a pure, separable state) to $(d-1)/d$ (for a maximally mixed state of dimension $d$). In this context, a non-zero $S_{l}(t)$ quantifies the degree to which the atom has become entangled with the electromagnetic modes, resulting in a mixed reduced state.
	
	A comparative analysis of $\mathcal{F}(t)$ and $S_{l}(t)$ reveals a strong anti-correlation between the accuracy of the semiclassical approximation and the growth of quantum correlations (see Fig.~\ref{fig_04_I}). When $S_{l}(t)$ is low, indicating that the atom-field state is nearly separable, the fidelity $\mathcal{F}(t)$ approaches unity. This confirms that the semiclassical approximation is valid in regimes where the system behaves predominantly as a factorizable mean field. Conversely, when $S_{l}(t)$ rises, indicating strong atom-field entanglement, $\mathcal{F}(t)$  drops significantly. The oscillations in fidelity correspond directly to the periodic entanglement and disentanglement cycles characteristic of Jaynes-Cummings dynamics.
	
	This correlation highlights the fundamental limitation of the semiclassical approach: it relies on a mean-field factorization that neglects quantum correlations. Consequently, the approximation inevitably loses accuracy during time intervals where the back-action of the fields on the atom generates significant entanglement, rendering the atomic state mixed rather than pure.
	
	\section{Approximate solution for the full Hamiltonian}\label{sec 5}
	We have solved the semiclassical Hamiltonian and examined its accuracy in reproducing the dynamics of the fully quantized two-mode JCM. Having identified its regime of validity, we are now in a position to use it as a reference for constructing an approximate solution.
	Recalling the decomposition in Eq~\eqref{eq:decomposition}, the interaction time-evolution operator factorizes as $\hat{U}_I(t)=\hat{U}_{\mathrm{sc,I}}(t) \hat{U}_I^{(1)}(t)$, with $\hat{U}_{\mathrm{sc,I}}(t)$  given by Eq.~\eqref{eq:U_sc}, and the evolution operator $\hat{U}_I^{(1)}(t)$ satisfies 
	\begin{equation}
		i\partial_t\hat{U}_I^{(1)}(t) =
		\left( \hat{U}_{\mathrm{sc,I}}^{\dagger}(t)\hat{V}^{(1)}(t) \hat{U}_{\mathrm{sc,I}}(t)\right)\hat{U}_I^{(1)}(t) = \hat{H}_I^{(1)}(t)\hat{U}_I^{(1)}(t).
	\end{equation}
	The Hamiltonian $\hat{H}_I^{(1)}(t)$ encodes all dynamics that are not captured by the semiclassical reference, including quantum fluctuations and back-action between the atom and the fields.
	Since $\hat{U}_{\mathrm{sc,I}}(t)$ acts only on the atomic degrees of freedom, we can define time-dependent operators $\hat{\sigma}_{\pm}^{\mathrm{(sc)}}(t) $ that evolve under the semiclassical reference frame as:
	\begin{eqnarray}\label{eq:simapm}
		\hat{\sigma}_{-}^{\mathrm{(sc)}}(t) & = & \hat{U}_{\mathrm{sc,I}}^{\dagger}(t)\hat{\sigma}_{-}\hat{U}_{\mathrm{sc,I}}(t) = \phi_1(t)\hat{\sigma}_{+}+\phi_2(t)\hat{\sigma}_{z}+\phi_3(t)\hat{\sigma}_{-}, \nonumber \\
		\hat{\sigma}_{+}^{\mathrm{(sc)}}(t) & = & \hat{U}_{\mathrm{sc,I}}^{\dagger}(t)\hat{\sigma}_{+}\hat{U}_{\mathrm{sc,I}}(t)  =   \phi_3^{*}(t)\hat{\sigma}_{+}+\phi_2^{*}(t)\hat{\sigma}_{z}+\phi_1^{*}(t)\hat{\sigma}_{-} ,  
	\end{eqnarray}
	with known functions $\phi_i(t)$ given by
	\begin{eqnarray}
		\phi_1(t)&=& -\beta_+^2(t)e^{-2\beta_z(t)},\nonumber
		\\
		\phi_2(t)&=&-\beta_+(t)(1+\beta_+(t)\beta_-(t)),\nonumber
		\\    \phi_3(t)&=&e^{2\beta_z(t)}\Big(1+\beta_+(t)\beta_-(t)(2+\beta_+(t)\beta_-(t))\Big).
	\end{eqnarray}
	Using this notation, the Hamiltonian $\hat{H}_{\mathrm{I}}^{(1)}(t)$ can be written as
	\begin{equation}\label{eq:H_dressed}
		\hat{H}_{\mathrm{I}}^{(1)}(t) = \sum_{i=1}^2 g_ie^{\mi \Delta_i t}\hat a_i\hat{\sigma}_+^{(\mathrm{sc})}(t)-\Omega_{\mathrm{sc}}(t)\hat{\sigma}_+^{(\mathrm{sc})}(t)+\mathrm{h.c.}
	\end{equation}
	
	Notice that the Hamiltonian $\hat{H}_{\mathrm{I}}^{(1)}(t) $ naturally separates into two distinct classes of terms, those proportional to $\hat\sigma_z$ and those proportional to $\hat{\sigma}_{\pm}.$ This observation suggests the following natural decomposition
	\[ \hat{H}_{I}^{(1)}(t) = \hat{H}_{0}^{(2)}(t) + \hat{V}^{(2)}(t) \]
	where
	\begin{equation}
		\label{H_0_2}
		\hat{H}_{0}^{(2)}(t) = \sum_{i=1}^2 g_ie^{\mi \Delta_i t}\phi_2^{*}(t)\hat{a}_i\hat{\sigma}_z -\Omega_{\mathrm{sc}}(t)\phi_2^{*}(t)\hat{\sigma}_z+ \mathrm{h.c.}
	\end{equation}
	and
	\begin{equation}
		\hat{V}^{(2)}(t) = \sum_{i=1}^2 g_ie^{\mi \Delta_i t}\hat{a}_i(\phi_3^{*}(t)\hat{\sigma}_{+}+\phi_1^{*}(t)\hat{\sigma}_{-})-\Omega_{\mathrm{sc}}(t)(\phi_3^{*}(t)\hat{\sigma}_{+}+\phi_1^{*}(t)\hat{\sigma}_{-}) +\mathrm{h.c.} 
	\end{equation}
	The time evolution operator $\hat{U}_{I}^{(1)}(t)$ can then be factorized as follows:
	\begin{equation}
		\hat{U}_{I}^{(1)}(t) = \hat{U}_{0}^{(2)}(t)\hat{U}_{I}^{(2)}(t).
	\end{equation}
	The factor $\hat{U}_{0}^{(2)}(t)$ can be written in terms of Glauber displacement operators 
	\begin{equation}\label{eq:Ui1}
		\hat{U}_{0}^{(2)}(t) = e^{\frac{1}{2}(\vert\epsilon_1(t)\vert^2+\vert\epsilon_2(t)\vert^2+2\epsilon_6(t))}e^{\epsilon_5(t)\hat{\sigma}_z}\hat D_{\hat{a}_1}(\epsilon_1(t)\hat{\sigma}_z)\hat D_{\hat{a}_2}(\epsilon_2(t)\hat{\sigma}_z),
	\end{equation}
	with complex, time-dependent functions $\epsilon_i(t)$ given in Appendix~\ref{Appendix B}.  
	\\
	The remaining factor $\hat{U}_{I}^{(2)}(t)$ satisfies the interaction Schrödinger equation
	\begin{equation}
		\label{eq_U_I_2}
		i\partial_t \hat{U}_{I}^{(2)}(t) = \left( \hat{U}_{0}^{(2)\dagger}(t) \hat{V}^{(2)}(t) \hat{U}_{0}^{(2)}(t) \right) \hat{U}_{I}^{(2)}(t) = \hat{H}_{I}^{(2)}(t) \hat{U}_{I}^{(2)}(t).
	\end{equation} 
	Transforming the atomic and field operators yields
	\begin{eqnarray*}
		\hat{U}_{0}^{(2)\dagger}(t)\hat{a}_i \hat{U}_{0}^{(2)}(t)\equiv \hat{a}_i^{(2)}(t)&=&\hat{a}_i+\epsilon_i(t)\hat{\sigma}_z\\
		\hat{U}_{0}^{(2)\dagger}(t)\hat{\sigma}_+\hat{U}_{0}^{(2)}(t)\equiv \hat{\sigma}_+^{(2)}(t)&=&e^{-2\epsilon_5(t)}e^{-2\epsilon_1(t)\hat{a}^{\dagger}_1}e^{2\epsilon_1^*(t)\hat{a}_1}e^{-2\epsilon_2(t)\hat{a}^{\dagger}_2}e^{2\epsilon_2^*(t)\hat{a}_2}e^{-2(\vert\epsilon_1(t)\vert^2+\vert\epsilon_2(t)\vert^2)\hat{\sigma}_z}\hat{\sigma}_+\\
		\hat{U}_{0}^{(2)\dagger}(t)\hat{\sigma}_-\hat{U}_{0}^{(2)}(t)\equiv \hat{\sigma}_-^{(2)}(t)&=&e^{2\epsilon_5(t)}e^{2\epsilon_1(t)\hat{a}^{\dagger}_1}e^{-2\epsilon_1^*(t)\hat{a}_1}e^{2\epsilon_2(t)\hat{a}^{\dagger}_2}e^{-2\epsilon_2^*(t)\hat{a}_2}e^{2(\vert\epsilon_1(t)\vert^2+\vert\epsilon_2(t)\vert^2)\hat{\sigma}_z}\hat{\sigma}_-
	\end{eqnarray*}
	
	Notice that the transformed atomic operators are products of exponentials containing field operators, multiplied by the atomic operators $\hat{\sigma}_{\pm}$. This structure makes $\hat{H}_{I}^{(2)}(t)$ highly cumbersome. To obtain a tractable Hamiltonian, we approximate these transformed atomic operators by their expectation value with respect to the most general  initial atomic state and coherent states for the fields. 
	
	\begin{equation}\label{initial state}
		\ket{\Psi(0)} = (C_e \ket{e} + C_g \ket{g}) \otimes \ket{\alpha_1, \alpha_2}.
	\end{equation}
	i.e.,
	\begin{equation}
		\hat{\sigma}_{\pm}^{(2)}(t) \approx \langle \hat{\sigma}_{\pm}^{(2)}(t)\rangle   
	\end{equation}
	where
	\begin{equation}
		\langle \hat{\sigma}_{\pm}^{(2)}(t) \rangle = C_e C_g \exp\!\left\{ \mp 4\left[ \frac{1}{2}\big( \epsilon_5(t) \pm (|\epsilon_1(t)|^2 + |\epsilon_2(t)|^2) \big) + i \Im(\alpha_1 \epsilon_1(t) + \alpha_2 \epsilon_2(t)) \right] \right\}.
	\end{equation}
	Under this approximation, the effective interaction Hamiltonian becomes
	\begin{align}
		\label{effective_H}
		\hat{H}_{I}^{(2)}(t) \simeq& \sum_{i=1}^2 g_ie^{i\Delta_i t}\Big[(\hat{a}_i+\epsilon_i(t) \hat{\sigma}_z(t))(\phi_3^{*}(t)\langle \hat{\sigma}_+^{(2)}(t)\rangle +\phi_1^{*}(t)\langle \hat{\sigma}_-^{(2)}(t)\rangle)\Big]\nonumber \\
		-& \Omega_{\mathrm{sc}}(t)(\phi_3^{*}(t)\langle \hat{\sigma}_+^{(2)}(t)\rangle +\phi_1^{*}(t)\langle \hat{\sigma}_-^{(2)}(t)\rangle)+ \mathrm{h. c.}
	\end{align}
	The corresponding time-evolution operator $\hat{U}_{I}^{(2)}(t)$ can also be written in terms of displacement operators:
	\begin{equation*}
		\hat{U}_{\mathrm{I}}^{(2)}(t) = e^{\frac{1}{2}(\vert\gamma_1(t)\vert^2+(\vert\gamma_2(t)\vert^2+2\gamma_6(t))}e^{\gamma_5(t)\sigma_z}\hat D_{a_1}(\gamma_1(t))\hat D_{a_2}(\gamma_2(t)).
	\end{equation*}
	with complex, time-dependent functions $\gamma_i(t)$ (see Appendix~\ref{Appendix B}).
	
	Finally, the full time-evolution operator is given by the product:
	\begin{equation}\label{full evolution operator}
		\hat{U}(t) = \hat{U}_0(t) \hat{U}_{\mathrm{sc}}(t) \hat{U}_{0}^{(2)}(t) \hat{U}_I^{(2)}(t). 
	\end{equation}
	Each factor corresponds to a distinct physical process: $\hat{U}_0(t)$ accounts for free evolution, $\hat{U}_{\mathrm{sc,I}}(t)$ describes the semiclassical atomic dynamics driven by the mean fields, and $\hat{U}_{0}^{(2)}(t)$ generates conditional displacements—the field displacement depends on whether the atom is in $|e\rangle$ or $|g\rangle$, encoding the back-action that leads to entanglement. Finally, $\hat{U}_{I}^{(2)}(t)$ captures the remaining driven field evolution, which includes an effective drive on the fields and a $\hat{\sigma}_z$ dependent feedback on the atom arising from the conditional displacements. 
	
	The time-evolved state follows from
	\[\hat U(t)\ket{\Psi(0)}=\ket{\Psi(t)}.\]
	with the initial state given by Eq.~\eqref{initial state}. 
	After carrying out the products of the evolution operators in Eq.~\eqref{full evolution operator}, the state takes the form
	\begin{equation}\label{eq: full vector state}
		\begin{aligned}
			\ket{\Psi(t)}=&F_0(t)\Big[C_eF_1(t)\Big( \beta_-(t)e^{\mi\omega_0t/2}\ket{g}+(1+\beta_+(t)\beta_-(t))e^{-\mi\omega_0t/2}\ket{e}\Big)\otimes\ket{\Gamma_1^{+}(t),\Gamma_2^{+}(t)}\\
			+& C_gF_2(t)\Big(e^{\mi \omega_0t/2}\ket{g}+\beta_+(t)e^{-\mi\omega_0t/2}\ket{e}\Big)\otimes\ket{\Gamma_1^{-}(t),\Gamma_2^{-}(t)}
			\Big],
		\end{aligned}
	\end{equation}
	where
	\begin{eqnarray*}
		F_0(t)&=&\mathrm{exp}\Big[\frac{1}{2}\left(\vert\epsilon_1(t)\vert^2+\vert\epsilon_2(t)\vert^2+\vert\gamma_1(t)\vert^2+\vert\gamma_2(t)\vert^2+2(\epsilon_6(t)+\gamma_5(t)+i\Im[\alpha_1(t)\gamma_1(t)+\alpha_2(t)\gamma_2(t)])\right)\Big],\\
		F_1(t)&=&\mathrm{exp}\Big[\beta_z(t)+\epsilon_5(t)+\mi\Im[\epsilon_1(t)(\alpha_1(t)+\gamma_1(t))^*+\epsilon_2(t)(\alpha_2(t)+\gamma_2(t))^*]\Big],\\
		F_2(t)&=&\mathrm{exp}\Big[-\left(\beta_z(t)+\epsilon_5(t)+\mi\Im[\epsilon_1(t)(\alpha_1(t)+\gamma_1(t))^*+\epsilon_2(t)(\alpha_2(t)+\gamma_2(t))^*]\right)\Big]    
	\end{eqnarray*}
	and the conditional field displacements are given by
	\begin{equation}
		\Gamma_{j}^{\pm}(t) = e^{-\mi\omega_j t} (\alpha_j + \gamma_j(t) \pm \epsilon_j(t)), \qquad j = 1,2.
	\end{equation}
	The structure of Eq.~\eqref{eq: full vector state} reveals the key feature of the approximation: the field states $\ket{\Gamma_1^{\pm}(t), \Gamma_2^{\pm}(t)}$ are displaced by $\pm \epsilon_j(t)$ depending on the atomic component, encoding the conditional dynamics that generate entanglement.
	
	\subsection{Results}
	
	The average photon number in mode $i$ is given by
	\begin{equation}
		\langle \hat{n}_i(t) \rangle = |\alpha_i|^2 + |\epsilon_i(t)|^2 + |\gamma_i(t)|^2 
		+ 2\Big[ \alpha_i \Re[\gamma_i(t)] + (|C_e|^2 - |C_g|^2) \Re[\epsilon_i(t)^* (\alpha_i + \gamma_i(t))] \Big].
	\end{equation}
	
	The first term, $|\alpha_i|^2$, corresponds to the classical coherent amplitude of the initial field. The terms $|\epsilon_i(t)|^2$ and $|\gamma_i(t)|^2$ arise from the conditional displacements and the driven evolution, respectively. The interference term $\alpha_i \Re[\gamma_i(t)]$ represents the mixing between the initial coherent state and the additional displacement. 
	Most importantly, the last term proportional to $|C_e|^2 - |C_g|^2$ encodes the back-action of the atom on the field: the photon number depends on whether the atom is more likely in the excited or ground state, a purely quantum feature absent in pure semiclassical treatments where the fields are unaffected by the atomic state.
	
	On the other hand, using the wavefunction given by Eq.~\eqref{eq: full vector state}, we computed the temporal evolution of the atomic inversion. 
	\begin{equation}
		\begin{aligned}
			\langle\hat{\sigma}_z(t)\rangle=&\vert F_0(t)\vert^2\Bigg[\vert C_eF_1(t)\vert^2\Big(\vert1+\beta_+(t)\beta_-(t)\vert^2-\vert\beta_-(t)\vert^2\Big)+\vert C_g F_2(t)\vert^2\Big(\vert\beta_+(t)\vert^2-1\Big)\\
			+& 2\Re\Big[C_eC_g^*F_1(t)F_2^*(t)\Big(\beta_+^*(t)(1+\beta_+(t)\beta_-(t))-\beta_-(t)\Big)\langle\Gamma_1^-(t),\Gamma_2^-(t)\vert \Gamma_1^+(t),\Gamma_2^+(t)\rangle\Big]\Bigg]
		\end{aligned}
	\end{equation}
	where
	\begin{equation*}
		\langle\Gamma_1^-(t),\Gamma_2^-(t)\vert \Gamma_1^+(t),\Gamma_2^+(t)\rangle=\mathrm{exp}\Big[-\frac{1}{2}\sum_{i=1}^2 \vert \Gamma_i^-(t)\vert^2+ \vert \Gamma_i^+(t)\vert^2-2\Gamma_i^{-*}(t)\Gamma_i^{+}(t)\Big].
	\end{equation*}
	From the atomic inversion expression, we highlight the third term. It captures, through the field overlap\\ $\langle \Gamma_1^-(t),\Gamma_2^-(t) | \Gamma_1^+(t),\Gamma_2^+(t) \rangle$, the atom–field entanglement generated by the conditional displacements. This overlap involves time-displaced coherent states and, together with the prefactors $F_0(t)$, $F_1(t)$, and $F_2(t)$, encodes the quantum corrections introduced by our approximation. These functions carry information about the evolution of the fields that become entangled with the atomic state. In the limit where the quantum corrections are neglected — i.e., when the conditional displacements vanish and the overlap reduces to unity — the expression simplifies considerably. For the particular case where the atom starts in the excited state ($C_e = 1$, $C_g = 0$) and disregarding the corrections encoded in $F_0(t)$, $F_1(t)$, and $F_2(t)$, we recover the pure semiclassical result given in  Eq.~\eqref{eq wei-norman atom}.
	
	\begin{figure}[htbp]
		\centering
		\includegraphics[width = \linewidth]{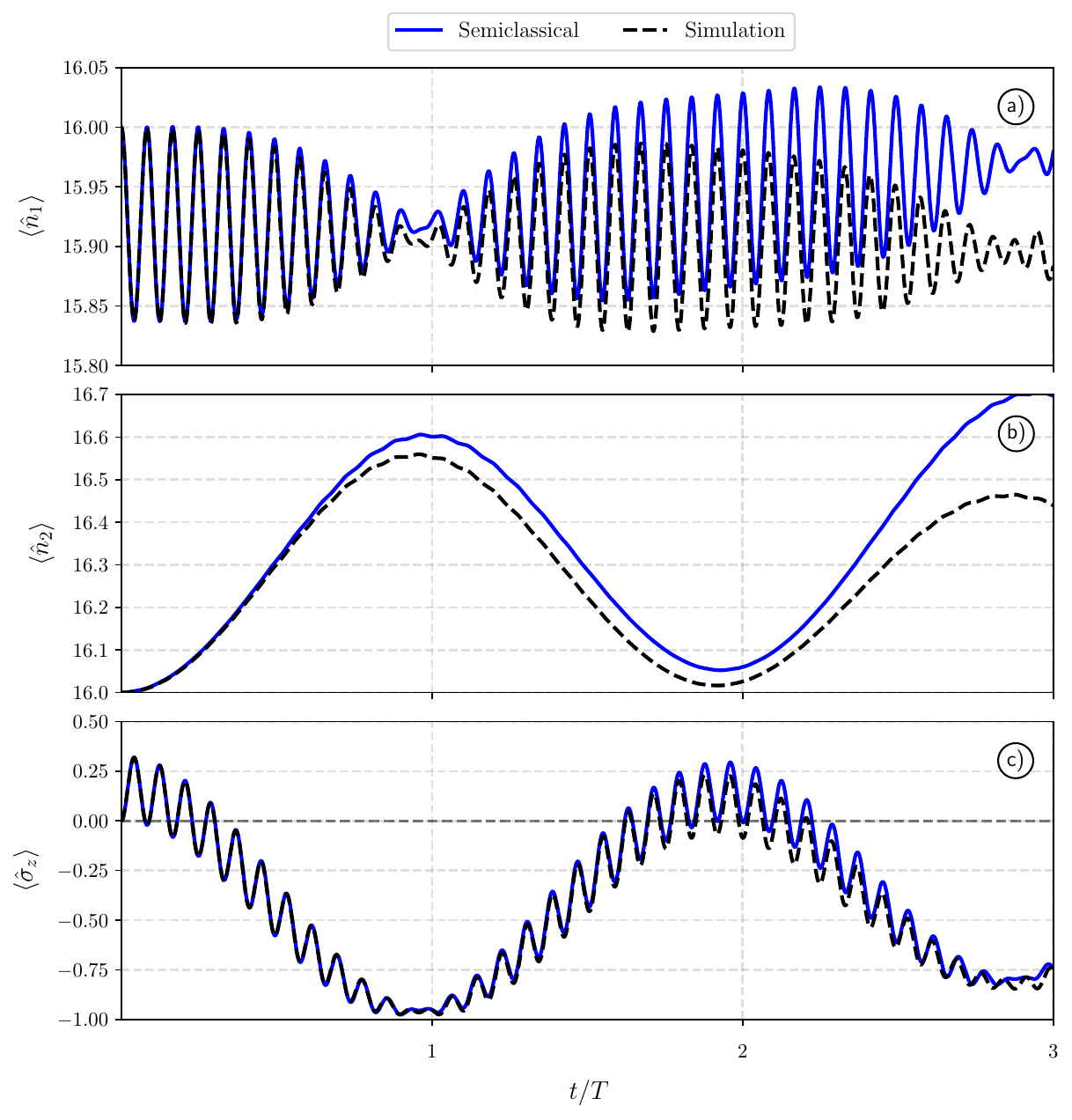}
		\caption{Time evolution of field observables and atomic inversion under the approximate full-dynamics scheme compared with exact numerical simulation. Panels (a) and (b) show the mean photon numbers $\langle \hat{n}_{1}(t) \rangle$ and $\langle \hat{n}_{2}(t) \rangle$, while panel (c) displays the atomic inversion $\langle \hat{\sigma}_{z}(t) \rangle$. The initial state is $|\Psi(0)\rangle = (C_e|e\rangle + C_g|g\rangle)\otimes|\alpha_1,\alpha_2\rangle$ with $\alpha_1 = \alpha_2 = 4$ and $C_e = C_g = 1/\sqrt{2}$. Parameters: $g_1 = 0.01\omega_2$, $g_{2} = g_{1}/5$, $\omega_{2} = 1$, $\omega_{1} = \omega_{2}/2$, $\omega_{0} = 0.98$. Despite $g_{1} > g_{2}$, mode 2 dominates the energy exchange due to near-resonance ($\Delta_{2} \ll \Delta_{1}$), producing large amplitude oscillations in $\langle \hat{n}_{2}(t) \rangle$ and $\langle \hat{\sigma}_{z}(t) \rangle$, while mode 1 contributes fast, small-amplitude modulations. The approximate solution (blue) reproduces the multi-timescale structure of the exact quantum dynamics (dashed black), with deviations emerging at long times due to accumulated atom--field correlations beyond the semiclassical factorization.}
		\label{fig_05}
	\end{figure}
	
	\begin{figure}[htbp]
		\centering
		\includegraphics[width = \linewidth]{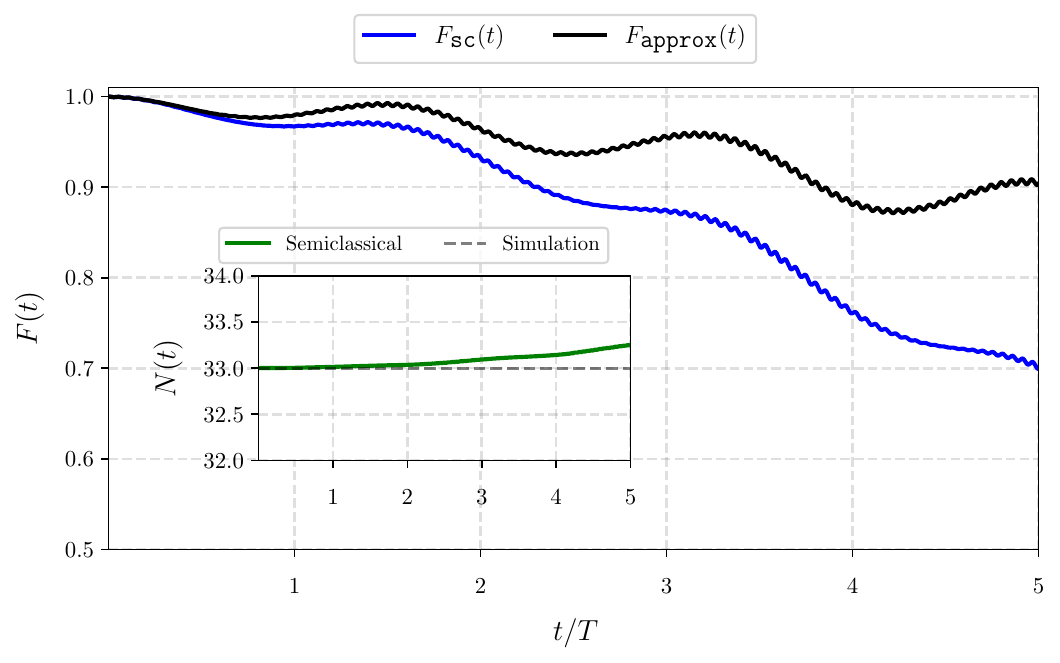}
		\caption{Fidelity between first approximation, Eq. (26), second approximation, Eq. (46), and simulation. In the inset, the total number of excitations in the second approximation and quantum simulation. $g_{1} = 0.01\omega_2$,  $\omega_2 = 1.0$, $\omega_{1} = \omega_{2} / 4$, $\omega_{0} = 0.98\omega_{2}$, $g_{2}^{(\texttt{opt})} = 0.00196\omega_2$, $\alpha_{1} = \alpha_2 = 4.0$, $C_{e} = 1.0$, $C_{g} = 0.0$.}
		\label{fig_06}
	\end{figure}
	
	In Fig.~\ref{fig_05}, we show in panel $a)$ the average value of the photon number operator for mode 1, in panel $b)$ the average value of the photon number operator for mode 2. In both cases, we compare our approximate results with those obtained by purely numerical means using the full Hamiltonian Eq.~\eqref{eq:JCM}. 
	The initial state of the system is $|\Psi(0)\rangle = (C_e |e\rangle + C_g |g\rangle)\otimes |\alpha_1,\alpha_2\rangle$ with $\alpha_1=\alpha_2=4$, $C_e=C_g$. In this calculation the coupling between mode 1 and the TLS is larger than the coupling between mode 2 and the TLS ($g_1 = 5 g_2$) however, mode two is near resonant with the TLS ($\Delta_2=0.02$) while mode 1 is far from resonance ($\Delta_1=0.48$) this combination of parameters is responsible for the fact that neither mode can be considered as a simple spectator.
	At the initial time $\langle \hat n_1 \rangle = \langle \hat n_2\rangle = 16$ and $\langle \hat\sigma_z \rangle =0$. In panel $a)$, we see a rapid exchange of excitations between mode 1 and the TLS with a period close to $2\pi/\Delta_1$. Superimposed on these rapid oscillations, there is an envelope with a much larger period of the order of $2\pi/\Delta_2$.  Since mode 1 is far from resonance, the amplitude of the oscillations is small, of the order of $0.15$ at the maximum. We can see that for long evolution times, the approximate result starts to deviate from the numerical simulation. This deviation is a consequence of the approximation we introduced when  replacing  time-evolved atomic operators by their mean values, $\hat{\sigma}_{\pm}^{(2)}(t) \approx \langle \hat{\sigma}_{\pm}^{(2)}(t)\rangle  $. This leads to an effective Hamiltonian (Eq.~\eqref{effective_H}) describing two driven fields, for which the total number of excitations is no longer a constant of the motion.
	
	We stress the accord in the qualitative agreement between our approximate results and the numerical. In panel $b)$ we see the evolution of $\langle n_2(t)\rangle.$ It consists of an oscillation with a period $2\pi/\Delta_2$ and a much larger amplitude than that of $\langle n_1(t)\rangle$, of the order of 0.6 at its maximum. At this time, the atomic inversion attains its minimum value and the TLS has reached its ground state. This is consistent with the fact that mode two is almost resonant with the TLS. In panel  $c)$ we see that the TLS couples effectively with mode 2 and presents an exchange of excitations with the same period of panel $b)$ on top of which there are rapid oscillations with much smaller amplitude whose frequency is that of panel $a)$. 
	
	Previously, we assessed the pure semiclassical approximation by computing the fidelity of the reduced atomic state. We now extend this analysis to the full quantum state, comparing both the pure semiclassical approach and our proposed approximation against the exact numerical solution of the full Hamiltonian. Specifically, we define $\mathcal{F}_{\mathrm{sc}}(t)$ as the fidelity between the semiclassical wavefunction from Eq.~\eqref{eq: atom vector state}  and the purely numerical solution of Eq.~\eqref{eq:JCM}. Similarly, $\mathcal{F}_{\mathrm{approx}}(t)$ is defined as the fidelity between the wavefunction obtained from our approximation in Eq.~\eqref{eq: full vector state} and the same numerical solution. Figure~\eqref{fig_06} compares the fidelity of the pure semiclassical approximation (blue) and the developed approximation (black).
	
	For the pure semiclassical approximation, the fidelity decays rapidly, dropping below $0.9$ after approximately $t=2T$ and below $0.8$ after $t=4T$, where $T = \pi/\Delta_2$. In contrast, the proposed approximation maintains significantly higher fidelity over a much longer timescale, remaining above $0.95$ for $t > 5T$. This demonstrates that the approximation reproduces the  quantum dynamics with significantly higher accuracy than the pure semiclassical approach over the timescales considered.
	
	Finally, the inset of Fig.~\ref{fig_06} shows the time-evolution of the average total excitation number $\hat{N} = \hat{n}_1 + \hat{n}_2 + \frac{1}{2}(\hat{\sigma}_z + 1),$ obtained from both wavefunction in~\eqref{eq: full vector state} (green) and the numerical simulation (black dotted line). As noted earlier, $\hat{N}$ commutes with the full Hamiltonian and is therefore a constant of the motion, in agreement with the numerical result. However, due to the approximations made in our scheme,
	$\hat{N}$ becomes time-dependent and is no longer conserved. Nevertheless, the approximate result remains very close to its initial value throughout the evolution, providing further evidence of the pertinence of our approximation.
	
	\section{Conclusions}
	\label{sec 6}
	
	In this work, we have developed an accurate approximate solution for a two-level system coupled to two off-resonant electromagnetic modes under the Jaynes–Cummings interaction. Our approach began with an exact decomposition of the Hamiltonian into two parts: a semiclassical component describing the atom driven by the mean fields of both modes, and a residual part containing the quantum fluctuations around this semiclassical reference.
	The semiclassical component was solved using two complementary methods. First, a first-order Magnus expansion provided analytical insight, revealing a field-field interference term oscillating at the beat frequency $|\omega_1-\omega_2|$ and yielding a design rule for maximum atomic coherence, $g_2^{\mathrm{opt}} = \pi\Delta_2/(8\alpha_2)$. Second, an exact solution via the Wei-Norman theorem served as a reference to validate the semiclassical dynamics.\\
	The residual part was treated through a sequence of unitary transformations that preserve the essential quantum features. By replacing the time-evolved operator $\hat{\sigma}_\pm^{(2)}(t)$ by their expectation values, we obtained an approximate Hamiltonian that captures conditional displacements and entanglement generation and whose time evolution operator could be obtained as a product of exponentials by means of the Wei-Norman theorem. \\
	The accuracy of the approximation was validated by comparing the atomic inversion, average photon number, and state fidelity with purely numerical simulations using the full two-mode JCM. The fidelity comparison (Fig.~\eqref{fig_06}) shows that the approximate wavefunction maintains high fidelity for significantly longer times than the pure semiclassical wavefunction, capturing quantum correlations that the latter misses entirely. The inset confirms that the approximation almost respects  the conservation of total excitation number, a fundamental physical constraint for the system.

	\begin{acknowledgments}
		J.R. and L.M.D. acknowledge partial support from project DGAPA UNAM IN 117925. 
		C.A.G-G acknowledges funding from Secretar\'ia de Ciencia, Humanidades, Tecnolog\'ia e Innovaci\'on (SECIHTI) Mexico, under grant No. CBF2023-2024-2888, and by DGAPA-PAPIIT-UNAM under grant No. IA104625. A.R.U. acknowledges Reyes Garc\'ia (C\'omputo-ICF) for maintaining the computational clusters, ICF-UNAM for the facilities provided during the research stay, and FCQeI-UAEM for partial support through teaching appointments.
	\end{acknowledgments}
	
	\newpage
	\appendix
	\section{ First order Magnus expansion}\label{Appendix A}
	The Magnus expansion \cite{Magnus1954} is a powerful technique for solving linear differential equations of the form
	\begin{equation}
		\frac{d}{dt}\boldsymbol{y}(t) = A(t) \boldsymbol{y}(t),
	\end{equation}
	where $A(t)$ is a time-dependent generator that does not necessarily commute with itself at different times, i.e., $[A(t_1), A(t_2)] \neq 0$. The solution is written as $\boldsymbol{y}(t) = \exp(\Omega(t)) \boldsymbol{y}(0)$, with $\Omega(t)$ expressed as an infinite series of nested commutators of $A(t)$:
	\begin{equation}
		\Omega(t) = \int_0^t A(t_1) dt_1 + \frac{1}{2} \int_0^t dt_1 \int_0^{t_1} dt_2 [A(t_1), A(t_2)] + \cdots
	\end{equation}
	When the time scale of variation of $A(t)$ is short compared to the evolution time or when the commutators are small, truncating the series at first order already provides an accurate approximation.
	This first-order Magnus approximation,
	\begin{equation}
		\label{magnus_1_order}
		\Omega_1(t) = \int_0^t A(\tau) d\tau,
	\end{equation}
	is valid when $\| \Omega_1(t) \| \ll 1$ or, more generally, when the higher-order commutators are negligible. 
	
	To apply this expansion method, we first express the system given by equations ~\eqref{eq:mf_s} and ~\eqref{eq:mf_w} in matrix form. Let $s(t) = s_{re}(t) + \mi s_{im}(t)$, where $s_{re}(t) = \mathrm{Re}[s(t)]$ and $s_{im}(t) \equiv \mathrm{Im}[s(t)]$. Next, we can define a real-valued vector
	\begin{equation}
		\boldsymbol{y}(t) = 
		\begin{bmatrix}
			W(t) \\ s_{re}(t) \\ s_{im}(t) 
		\end{bmatrix}.
	\end{equation}
	With this, we can rewrite the complex Rabi frequency, $\Omega_{\texttt{sc}}(t)$, using Euler's formula, $e^{\pm \mi \Delta_{k} t} = \cos(\Delta_{k} t) \pm \mi \sin(\Delta_{k} t)$, obtaining the relations
	\begin{equation}
		\begin{aligned}
			\Omega_{\texttt{sc}}(t) &= U(t) + \mi V(t), \\
			\Omega^{*}_{\texttt{sc}}(t) &= U(t) - \mi V(t),
		\end{aligned}
	\end{equation}
	where
	\begin{equation}
		U(t) = \sum_{i = 1}^{2} g_{i}\alpha_{i}\cos\left(\Delta_{i} t\right),\qquad
		V(t) = \sum_{i = 1}^{2} g_{i}\alpha_{i}\sin\left(\Delta_{i} t\right).
	\end{equation}
	We then substitute these results into Eqs.~\eqref{eq:mf_s} and \eqref{eq:mf_w}, where a real and imaginary separation yields
	\begin{equation}
		\begin{gathered}
			\dot{W}(t) = 4V(t) s_{re}(t) - 4U(t) s_{im}(t), \\
			\dot{s}_{re}(t) = -V(t) W(t),\qquad
			\dot{s}_{im}(t) = U(t) W(t).
		\end{gathered}
	\end{equation}
	We can notice that these differential equations can be written in matrix form as
	\begin{equation}
		\frac{d}{dt} \boldsymbol{y}(t) = A(t) \, \boldsymbol{y}(t),
	\end{equation}
	where the time-dependent matrix $A(t)$ is given by
	\begin{equation}\label{eq:Amatrix}
		A(t) = 
		\begin{bmatrix}
			0 & 4V(t) & -4U(t) \\
			-V(t) & 0 & 0 \\
			U(t) & 0 & 0
		\end{bmatrix}. 
	\end{equation}
	
	The formal solution to $\dot{\boldsymbol{y}}(t) = A(t) \boldsymbol{y}(t)$ is $\boldsymbol{y}(t) = \exp(\Omega(t)) \boldsymbol{y}(0)$, where $\Omega(t)$ is the proper Magnus expansion of the matrix $A(t)$. A first-order expansion around $A(0)$ yields
	\begin{equation}
		\boldsymbol{y}(t) \approx \exp(\Omega_{1}(t)) \, \boldsymbol{y}(0).
	\end{equation}
	
	with $ \Omega_{1}(t)$ given be equation ~\eqref{magnus_1_order}. We then proceed to integrate this expression. Let us define the integrals of the driving functions, $U(t)$ and $V(t)$, as
	\begin{align}
		\nonumber I_{V}(t) &= \int_{0}^{t} V(\tau) \, d\tau\\
		&= \sum\limits_{i = 1}^{2} g_{i}\alpha_{i}\frac{1 - \cos(\Delta_{i}t)}{\Delta_{i}} \equiv a(t), \label{eq:aInt} \\
		\nonumber I_{U}(t) &= \int_{0}^{t} U(\tau) \, d\tau\\
		&= \sum\limits_{i = 1}^{2} g_{i}\alpha_{i}\frac{\sin(\Delta_{i} t)}{\Delta_{i}} \equiv b(t). \label{eq:bInt}
	\end{align}
	Where the first-order Magnus term, named $M$, is
	\begin{equation}
		\begin{aligned}
			\Omega_{1}(t) &= \int_{0}^{t} A(\tau) d\tau\\
			& =
			\begin{bmatrix}
				0 & 4a(t) & -4b(t) \\
				-a(t) & 0 & 0 \\
				b(t) & 0 & 0
			\end{bmatrix} \equiv M.
		\end{aligned}
	\end{equation}
	
	It is worth noting that $M$ is an element of the $\mathfrak{sl}(3, \mathbb{R})$ Lie algebra. Since $\mathrm{tr}(M) = 0$, for each fixed $t$ we have $\mathrm{det}(\exp(M)) = \exp(\mathrm{tr}(M)) = 1$. Its spectral properties give that the minimal polynomial is $\lambda(\lambda^{2} + 4\omega^{2})$, for $\omega^{2} = a(t)^{2} + b(t)^{2}$, which coincides with its Casimir invariant $\mathcal{C}$. It thus define the relation $M^{3} = - 2\omega^{2}M$. The Caley-Hamilton series reduction yields
	\begin{equation}
		\exp(M) = \mathbb{I} + \frac{\sin \omega}{\omega} M + \frac{1 - \cos \omega}{\omega^2} M^2, \label{eq:expMformula}
	\end{equation}
	where $\mathbb{I}$ is the $3 \times 3$ identity matrix. It thus suffice to obtain $M^{2}$ as
	\begin{equation}
		M^{2} = 
		\begin{bmatrix}
			-\omega^2 & 0 & 0 \\
			0 & -4a^2 & 4ab \\
			0 & 4ab & -4b^2
		\end{bmatrix}.
	\end{equation}
	Finally, substituting back into Eq.~\eqref{eq:expMformula} and simplifying we obtain
	\begin{equation}
		\exp(M) = 
		\begin{bmatrix}
			\cos \omega & \dfrac{4a \sin \omega}{\omega} & -\dfrac{4b \sin \omega}{\omega} \\[10pt]
			-\dfrac{a \sin \omega}{\omega} & 1 - \dfrac{4a^2 (1 - \cos \omega)}{\omega^2} & \dfrac{4ab (1 - \cos \omega)}{\omega^2} \\[10pt]
			\dfrac{b \sin \omega}{\omega} & \dfrac{4ab (1 - \cos \omega)}{\omega^2} & 1 - \dfrac{4b^2 (1 - \cos \omega)}{\omega^2}
		\end{bmatrix}. \label{eq:expMexplicit}
	\end{equation}
	The one-parameter subgroup $\exp(sM)$ generated by $M$ at fixed $t$ belongs to the $\mathrm{SL}(3, \mathbb{R})$ group. And because $\mathfrak{sl}(3, \mathbb{R})$ is a compact algebra, it results that its group is conjugate to an $\mathrm{SO}(3)$ rotation subgroup inside $\mathrm{SL}(3, \mathbb{R})$. Since $\omega^{2}$ it's composed of explicit oscillatory functions, $M$ sweeps through the algebra, producing a curve $t\mapsto M(t)$ in $\mathfrak{so}(3)$. We thus can approximate the real vector $\psi$ at time $t$ by 
	\begin{equation}
		\boldsymbol{y}(t) \approx \exp(M) \, \boldsymbol{y}(0).
	\end{equation}
	For the common initial condition where the system starts in the excited state
	\begin{equation}
		\boldsymbol{y}(0) = 
		\begin{bmatrix} 1 \\ 0 \\ 0 \end{bmatrix},
	\end{equation}
	where we can see that the solution is just the action of $\exp(M)$ on the single non-vanishing element, returning the quantities
	\begin{gather}
		W(t) \approx \cos \omega(t), \label{eq:solW} \\
		s_{re}(t) \approx -a(t) \frac{\sin \omega(t)}{\omega(t)},\qquad
		s_{im}(t) \approx -\frac{b(t)}{a(t)} s_{re}(t), \label{eq:solS}
	\end{gather}
	with $\omega(t)$ defined above via $a(t)$, $b(t)$ in Eqs.~\eqref{eq:aInt} and \eqref{eq:bInt}.
	The first-order Magnus expansion is valid only when the generator $M$ oscillates slowly in the Lie algebra, i.e. when the reference frequency $\omega_{0}$ is close to one of the driving frequencies $\omega_{i}$, or varies rapidly compared to the system's evolution timescale i.e. $\left\vert \tfrac{g_{i} \alpha_{i}}{\Delta_{i}} \right\vert \ll 1$. Otherwise, the effective rotation generated by $\exp(M)$ accumulates destructively, and it must include higher-order commutators to capture the dynamics.
	
	\section{ Building time-evolution operators via application of Wei-Norman theorem}\label{Appendix B}
	The Wei-Norman theorem provides a method to express the time-evolution
	operator of a quantum system whose Hamiltonian can be written as a
	linear combination of operators that close under commutation, i.e.,
	they form a finite-dimensional Lie algebra \cite{WeiNorman1964}. Let the Hamiltonian be
	written as
	\begin{equation}
		\hat H(t)=\sum_{k=1}^{n} h_k(t) \hat A_k ,
	\end{equation}
	where the operators $\hat A_k$ satisfy the commutation relations
	\begin{equation}
		[\hat A_i,\hat A_j]=\sum_{k=1}^{n} c_{ij}^{k} \hat A_k ,
	\end{equation}
	with $c_{ij}^{k}$ the structure constants of the algebra. The time
	evolution operator $\hat U(t)$ is defined by the Schrödinger equation
	\begin{equation}
		i\partial_t \hat U(t)
		=
		\hat H(t)\hat U(t), \qquad \hat{U}(t_0)=\cal{I}.
	\end{equation}
	Under these conditions, the Wei--Norman theorem states that the
	time-evolution operator can be written in the factorized form
	\begin{equation}
		\hat U(t)=\prod_{k=1}^{n} e^{x_k(t) \hat A_k},
	\end{equation}
	where $x_k(t)$ are time-dependent functions determined by substituting
	this ansatz into the Schrödinger equation, which leads to a system of
	coupled differential equations for the parameters $x_k(t)$.
	\begin{itemize}
		
		\item \textbf{Time-evolution operator $\mathbf{\hat U_{\mathrm{sc,I}}(t)}$.}
		The set of operators appearing in the semiclassical interaction Hamiltonian
		(Eq.~\eqref{eq:H_sc}) is closed under commutation. Consequently, the
		time-evolution operator can be written exactly as a product of
		exponentials.
		\begin{equation*}
			\hat U_{\mathrm{sc,I}}(t)
			=
			e^{\beta_{+}(t)\hat{\sigma}_{+}}
			e^{\beta_{-}(t)\hat{\sigma}_{-}}
			e^{\beta_{z}(t)\hat{\sigma}_{z}},
		\end{equation*}
		where $\beta_i(t)$ are complex time-dependent functions.
		
		Substituting this ansatz into the Schr\"odinger equation
		\begin{equation*}
			i\partial_t \hat U_{\mathrm{sc,I}}(t)
			=
			\hat H_{\mathrm{sc,I}}(t)\hat U_{\mathrm{sc,I}}(t),
		\end{equation*}
		one obtains a system of ordinary differential equations for the
		coefficients $\beta_i(t)$,
		\begin{eqnarray}\label{eq:ode}
			\dot{\beta}_z &=& i \beta_{+}\Omega_{\mathrm{sc}}^*(t), \\
			\dot{\beta}_{+} &=& -i\left(\Omega_{\mathrm{sc}}(t) - \beta_{+}^2 \Omega_{\mathrm{sc}}^*(t) \right), \\
			\dot{\beta}_{-} &=& -i(1+2\beta_{+}\beta_{-})\Omega_{\mathrm{sc}}^*(t).
		\end{eqnarray}
		\item \textbf{Time-evolution operator $\mathbf{\hat U_{0}^{(2)}(t)}$.}  The time-evolution operator associated to $\hat{H}_{0}^{(2)}(t)$ (Eq. ~\eqref{H_0_2})  can be written exactly as:
		\begin{equation*}
			\hat U_0^{(2)}(t)=e^{\epsilon_1(t)\hat{a}_1^{\dagger}\hat\sigma_z}e^{\epsilon_2(t)\hat{a}_2^{\dagger}\hat\sigma_z}e^{\epsilon_3(t)\hat{a}_1\hat\sigma_z}e^{\epsilon_4(t)\hat{a}_2\hat\sigma_z}e^{\epsilon_5(t)\hat\sigma_z}e^{\epsilon_6(t)},
		\end{equation*}
		$\hat U_0^{(2)}(t)$ is a solution of
		\[
		i\partial_t \hat U_0^{(2)}(t)
		=
		\hat H_{0}^{(2)}(t)\hat U_0^{(2)}(t).
		\]
		Therefore, $\epsilon_i(t)$ complex functions can be obtained by solving the following system of ODEs.
		\begin{align}
			\dot{\epsilon}_1(t) &= -ig_1\phi_2(t)e^{-i\Delta_1t}, \\
			\dot{\epsilon}_2(t) &= -ig_2\phi_2(t)e^{-i\Delta_2t}, \\
			\dot{\epsilon}_3(t) &= -ig_1\phi_2^*(t)e^{i\Delta_1t},\\
			\dot{\epsilon}_4(t) &= -ig_2\phi_2^*(t)e^{i\Delta_2t},\\
			\dot{\epsilon}_5(t) &= i(\Omega_{sc}(t)\phi_2^*(t)+\Omega_{sc}^*(t)\phi_2(t)),\\
			\dot{\epsilon}_6(t) &=-i(g_1\phi_2^*(t)e^{i\Delta_1t}\epsilon_1(t)+g_2\phi_2^*(t)e^{i\Delta_2t}\epsilon_2(t)).
		\end{align}
		We can notice the following relations $\epsilon_1(t)=-\epsilon_3^*(t)$ and $\epsilon_2(t)=-\epsilon_4^*(t)$. The time-dependent coefficients $\epsilon_i(t)$ are obtained by numerical integration.
		
		Taking those relations, $\hat U_0^{(2)}(t)$ can be written in terms of Glauber displacement operators \cite{glauber1963coherent} 
		\begin{equation*}
			\hat{U}_{0}^{(2)}(t) = e^{\frac{1}{2}(\vert\epsilon_1(t)\vert^2+(\vert\epsilon_2(t)\vert^2+2\epsilon_6(t))}e^{\epsilon_5(t)\hat\sigma_z}D_{\hat a_1}(\epsilon_1(t)\hat \sigma_z)D_{\hat a_2}(\epsilon_2(t)\hat\sigma_z).
		\end{equation*}
		\item \textbf{Time-evolution operator $\mathbf{\hat U_{\mathrm{I}}^{(2)}(t)}$.} Finally, the exact form of $U_{\mathrm{I}}^{(2)}(t)$ is given by
		\begin{equation*}
			U_\mathrm{I}^{(2)}(t)=e^{\gamma_1(t) \hat a^{\dagger}_1}e^{\gamma_2(t) \hat a^{\dagger}_2}e^{\gamma_3(t)\hat a_1}e^{\gamma_4(t)\hat a_2}e^{\gamma_5(t)\hat \sigma_z}e^{\gamma_6(t)}.
		\end{equation*}
		After substituting $ \hat U_\mathrm{I}^{(2)}$ into Schr\"odinger equation (Eq. ~\eqref{eq_U_I_2}), we obtain the corresponding ODEs system for $\gamma_i(t)$ functions.
		\begin{align}
			\dot{\gamma_1}(t) &= -ig_1e^{-i\Delta_1t}(\phi_1(t)\langle  \hat{\sigma}_+^{(2)}(t)\rangle+\phi_3(t)\langle \hat{\sigma}_-^{(2)}(t)\rangle), \\
			\dot{\gamma}_2(t) &= -ig_2e^{-i\Delta_2 t}(\phi_1(t)\langle  \hat{\sigma}_+^{(2)}(t)\rangle+\phi_3(t)\langle \hat{\sigma}_-^{(2)}(t)\rangle), \\
			\dot{\gamma}_3(t) &= -ig_1e^{i\Delta_1t}(\phi_3^*(t)\langle  \hat{\sigma}_+^{(2)}(t)\rangle+\phi_1^*(t)\langle \hat{\sigma}_-^{(2)}(t)\rangle),\\
			\dot{\gamma}_4(t) &= -ig_2e^{i\Delta_2t}(\phi_3^*(t)\langle  \hat{\sigma}_+^{(2)}(t)\rangle+\phi_1^*(t)\langle \hat{\sigma}_-^{(2)}(t)\rangle),\\
			\dot{\gamma}_5(t) &= -i\Big[(g_1e^{i\Delta_1t}\epsilon_1(t)+g_2e^{i\Delta_2t}\epsilon_2(t))(\phi_3^*(t)\langle  \hat{\sigma}_+^{(2)}(t)\rangle+\phi_1^*(t)\langle \hat{\sigma}_-^{(2)}(t)\rangle)+\mathrm{c.c.}\Big],\\
			\dot{\gamma}_6(t) &=-i\Big[(g_1e^{i\Delta_1t}\gamma_1(t)+g_2e^{i\Delta_2t}\gamma_2(t))(\phi_3^*(t)\langle  \hat{\sigma}_+^{(2)}(t)\rangle+\phi_1^*(t)\langle \hat{\sigma}_-^{(2)}(t)\rangle)\Big]\\
			&+i\Big[\Omega_{\mathrm{sc}}(t)(\phi_3^*(t)\langle  \hat{\sigma}_+^{(2)}(t)\rangle+\phi_1^*(t)\langle \hat{\sigma}_-^{(2)}(t)\rangle)+\mathrm{c.c.}\Big].
		\end{align}
		Similar to $\epsilon_i(t)$ system, we have the following relations: $\gamma_1(t)=-\gamma_3^*(t)$ and $\gamma_2(t)=-\gamma_4^*(t)$. These relations allow us to write $\hat U_\mathrm{I}^{(2)}(t)$ in terms of displacement operators
		\begin{equation*}
			\hat{U}_{\mathrm{I}}^{(2)}(t) = e^{\frac{1}{2}(\vert\gamma_1(t)\vert^2+(\vert\gamma_2(t)\vert^2+2\gamma_6(t))}e^{\gamma_5(t)\hat \sigma_z}D_{\hat a_1}(\gamma_1(t))D_{\hat a_2}(\gamma_2(t)).
		\end{equation*}
	\end{itemize}
	The time-dependent coefficients $\beta_i(t)$, $\epsilon_i(t)$, and $\gamma_i(t)$ are computed by numerical integration using Runge--Kutta (RK45) method.
	
	\bibliographystyle{apsrev4-2}
	\bibliography{refs}

@article{Kimble1998,
  author       = {Kimble, H. J.},
  title        = {Strong interactions of single atoms and photons in cavity {QED}},
  journal      = {Physica Scripta},
  volume       = {T76},
  pages        = {127--137},
  year         = {1998},
  doi          = {10.1238/Physica.Topical.076a00127}
}

@article{blas-moya,
    author = {BM Rodriguez-Lara and HM Moya-Cessa},
    title = {The exact solution of generalized Dicke models via Sussking-Glogower operators},
    journal = {Journal of Physics A },
    volume = {46},
    year = {2013},
    pages = {095301}
}

@article{moya,
    author = {HM Moya-Cessa},
    title = {Decoherence in atom-field interactions: a treatment using super operator techniques},
    journal = {Physics Reports} ,
    year = {2006},
    volume = {432},
    pages = {1-41}
}

@article{recamier2018,
    author = {C Gonz\'alez-Guti\'errez and O de los Santos-S\'anchez and R Rom\'an-Ancheyta and M Berrondo and J R\'ecamier },
    title = {Lie algebraic approach to a nonstationary atom-cavity system} ,
    journal = {Journal of the Optical Society of America B} ,
    year = {2018} ,
    volume = {35},
    number = {8},
    pages = {1979}
}

@article{vogel,
    author ={W Vogel and R.L. de Matos Filho} ,
    title ={Nonlinear Jaynes-Cummings dynamics of a trapped ion} ,
    journal ={Physical Review A} ,
    volume ={52},
    number={2},
    pages ={4214},
    year ={1995} 
}

@article{xiao,
    author ={Xiao-Hang Cheng and I Arrazola and JS Pedernales and L Lamata and Xi Chen and E Solano} ,
    title = {Nonlinear Quantum Rabi model in trapped ions},
    journal ={Physical Review A} ,
    volume = {97},
    pages ={023624},    
    year ={2018} 
}

@book{HarocheRaimond2006,
  author       = {Haroche, Serge and Raimond, Jean-Michel},
  title        = {Exploring the Quantum: Atoms, Cavities, and Photons},
  publisher    = {Oxford University Press},
  address      = {Oxford},
  year         = {2006},
  isbn         = {9780198509141}
}

@article{Leibfried2003,
  author       = {Leibfried, D. and Blatt, R. and Monroe, C. and Wineland, D. J.},
  title        = {Quantum dynamics of single trapped ions},
  journal      = {Reviews of Modern Physics},
  volume       = {75},
  pages        = {281--324},
  year         = {2003},
  doi          = {10.1103/RevModPhys.75.281}
}

@article{BlattRoos2012,
  author       = {Blatt, Rainer and Roos, Christian F.},
  title        = {Quantum simulations with trapped ions},
  journal      = {Nature Physics},
  volume       = {8},
  pages        = {277--284},
  year         = {2012},
  doi          = {10.1038/nphys2252}
}

@article{Blais2021,
  author       = {Blais, Alexandre and Grimsmo, Arne L. and Girvin, Steven M. and Oliver, William D.},
  title        = {Circuit quantum electrodynamics},
  journal      = {Reviews of Modern Physics},
  volume       = {93},
  pages        = {025005},
  year         = {2021},
  doi          = {10.1103/RevModPhys.93.025005}
}

@article{Reagor2016,
  author       = {Reagor, Matthew and Pfaff, Wolfgang and Axline, C. and Heeres, R. and Ofek, N. and Sliwa, K. and Holland, E. and Wang, C. and Blumoff, J. and Chou, K. and Hatridge, M. and Frunzio, L. and Schoelkopf, R. J. and Devoret, M. H.},
  title        = {Quantum memory with millisecond coherence in circuit {QED}},
  journal      = {Physical Review B},
  volume       = {94},
  pages        = {014506},
  year         = {2016},
  doi          = {10.1103/PhysRevB.94.014506}
}

@article{Baumann2010,
  author       = {Baumann, K. and Guerlin, C. and Brennecke, F. and Esslinger, T.},
  title        = {Dicke quantum phase transition with a superfluid gas in an optical cavity},
  journal      = {Nature},
  volume       = {464},
  pages        = {1301--1306},
  year         = {2010},
  doi          = {10.1038/nature09009}
}

@article{Gopalakrishnan2009,
  author       = {Gopalakrishnan, S. and Lev, B. L. and Goldbart, P. M.},
  title        = {Emergent crystallinity and frustration in multimode cavity {QED}},
  journal      = {Nature Physics},
  volume       = {5},
  pages        = {845--850},
  year         = {2009},
  doi          = {10.1038/nphys1323}
}

@article{Thompson2013,
  author       = {Thompson, J. D. and Tiecke, T. G. and Zibrov, A. S. and Vuleti{\'c}, V. and Lukin, M. D.},
  title        = {Coupling a single trapped atom to a nanoscale optical cavity},
  journal      = {Science},
  volume       = {340},
  pages        = {1202--1205},
  year         = {2013},
  doi          = {10.1126/science.1237125}
}

@article{Chang2014,
  author       = {Chang, D. E. and Vuleti{\'c}, V. and Lukin, M. D.},
  title        = {Quantum nonlinear optics --- photon by photon},
  journal      = {Nature Photonics},
  volume       = {8},
  pages        = {685--694},
  year         = {2014},
  doi          = {10.1038/nphoton.2014.192}
}

@article{gerry,
    author = { C. Gerry } ,
    title = {Dynamics of a two-photon, two-mode Jaynes-Cummings model},
    journal = {Journal of the Optical Society of America B} ,
    year = {1992},
    volume = {9},
    number = {2},
    pages = {290-297}
}

@article{castanos,
author = {L. O. Castaños-Cervantes },
title = {Coherent control of two Jaynes-Cummings cavities},
    journal = {Scientific Reports},
    year = {2024}, 
    volume = {14},
    pages = {3790}
}

@article{alscher,
    author = {Alscher, A and Grabert, H.} ,
    title = {Semiclassical dynamics of the Jaynes-Cummings model},
    journal = {The European Physical Journal D: Atomic, Molecular, Optical and Plasma Physics},
    year = {2001},
    volume = {14},
    number = {1},
    pages = {127-136}
    }

@article{babelon,
    author = {O. Babelon and L. Cantini and B. Deucot},
    title = {A semi-classical study of the Jaynes-Cummings model},
    journal = {J. Stat. Mech.},
    year = {2009},
    pages = {P07011}
}

@article{vidella,
    author = {J. Larson and T. Mavrogordatos and S. Parkins and A. Vidella-Barranco},
    title = {The Jaynes-Cummings model: 60 years and still counting},
    journal = {Journal of the Optical Society of America B},
    year = {2024},
    volume = {41},
    number = {8},
    pages = {JCM1-JCM4}
}

@article{WeiNorman1964,
  author       = {Wei, J. and Norman, E.},
  title        = {On global representations of the solutions of linear differential equations as a product of exponentials},
  journal      = {Proceedings of the American Mathematical Society},
  volume       = {15},
  pages        = {327--334},
  year         = {1964},
  doi          = {10.1090/S0002-9939-1964-0152064-7}
}

@article{Magnus1954,
  author       = {Magnus, Wilhelm},
  title        = {On the exponential solution of differential equations for a linear operator},
  journal      = {Communications on Pure and Applied Mathematics},
  volume       = {7},
  pages        = {649--673},
  year         = {1954},
  doi          = {10.1002/cpa.3160070404}
}

@article{glauber1963coherent,
  title={Coherent and incoherent states of the radiation field},
  author={Glauber, Roy J},
  journal={Physical Review},
  volume={131},
  number={6},
  pages={2766},
  year={1963},
  publisher={APS}
}

@article{jozsa1994fidelity,
  title={Fidelity for mixed quantum states},
  author={Jozsa, Richard},
  journal={Journal of modern optics},
  volume={41},
  number={12},
  pages={2315--2323},
  year={1994},
  publisher={Taylor \& Francis},
   doi={10.1080/09500349414552171}
}

@article{medina2020,
  title={Approximate evolution for a hybrid system—an optomechanical Jaynes-Cummings model},
  author={Medina-Dozal, Luis and Ramos-Prieto, Ir{\'a}n and R{\'e}camier, Jos{\'e}},
  journal={Entropy},
  volume={22},
  number={12},
  pages={1373},
  year={2020},
  publisher={MDPI}
}
\end{document}